\pdfoutput=1
\documentclass[ prl,twocolumn,superscriptaddress,amsmath,amssymb,aps,longbibliography,nofootinbib
]{revtex4-2}

\usepackage{graphicx}    
\usepackage{float}
\usepackage{placeins}   
\usepackage{bm}          
\usepackage{siunitx}     
\usepackage{hyperref}    
\usepackage{dcolumn}     
\usepackage{braket}      
\usepackage[T1]{fontenc}
\usepackage{lmodern} 
\usepackage{gensymb}
\usepackage{cleveref}    

\crefname{figure}{Fig.}{Figs.}
\Crefname{figure}{Figure}{Figures}
\crefname{equation}{Eq.}{Eqs.}
\Crefname{equation}{Equation}{Equations}
\crefname{table}{Table}{Tables}
\Crefname{table}{Table}{Tables}
\crefname{suppfigure}{Supplementary Fig.}{Supplementary Figs.}
\Crefname{suppfigure}{Supplementary Fig.}{Supplementary Figs.}
\crefname{suppequation}{Supplementary Eq.}{Supplementary Eqs.}
\Crefname{suppequation}{Supplementary Eq.}{Supplementary Eqs.}
\creflabelformat{suppequation}{#2(#1)#3}
\crefname{supptable}{Supplementary Table}{Supplementary Tables}
\Crefname{supptable}{Supplementary Table}{Supplementary Tables}
\crefname{suppnote}{Supplementary Note}{Supplementary Notes}
\Crefname{suppnote}{Supplementary Note}{Supplementary Notes}


\hypersetup{
  colorlinks=true,
  linkcolor=blue,
  citecolor=blue,
  urlcolor=blue
}

\begin{document}

\title{Unraveling square and kagome lattice  frustrated magnetism  by scalable tensor networks }

\author{Xing-Yu Zhang}
\affiliation{Department of Physics and Astronomy, Ghent University, Krijgslaan 281, 9000 Gent, Belgium}

\author{Jutho Haegeman}
\affiliation{Department of Physics and Astronomy, Ghent University, Krijgslaan 281, 9000 Gent, Belgium}

\author{Yuchi He}
\email{yuchi.he@ugent.be}
\affiliation{Department of Physics and Astronomy, Ghent University, Krijgslaan 281, 9000 Gent, Belgium}

\date{\today}

\begin{abstract}
We extend the scalability of variational infinite two-dimensional tensor network methods with application to investigate frustrated magnetism of the square-lattice $J_1$-$J_2$ model and kagome-lattice Heisenberg model. In the sign-problem-free limit of the square-lattice model, we achieve high ground-state energy precision comparable to quantum Monte Carlo methods. With no obstruction beyond this limit, our method obtains direct estimates of order parameters. For the square-lattice $J_1$-$J_2$ model, we determine isotropic plaquette order in the valence-bond solid region. For the model near $J_2/J_1=0.5$, and the kagome lattice Heisenberg model, our data indicate linear-dispersion gapless states.
To probe the candidate spin-liquid nature of the highly frustrated cases, we go beyond inferring the absence of symmetry-breaking orders and analyze susceptibilities that characterize gapless spin liquids. By estimating scaling dimensions, our data indicate a stable gapless spin liquid on the kagome lattice, and competing scenarios for the intermediate regime of the square-lattice model. Our developments include parallelization and randomized low-rank methods and largely expand the frontier of high-precision variational numerics for quantum phases of matter. 
\end{abstract}

\maketitle

\section{Introduction}
\label{sec:intro}
Spin models on lattices provide good descriptions of quantum magnets with localized moments. A fascinating possibility is that frustration suppresses conventional symmetry-breaking orders and gives rise to emergent fermions and ``light'', described by a U(1) gauge theory (quantum electrodynamics, QED)~\cite{PhysRevB.37.580}. The emergent gauge group can also be $\mathbb{Z}_2$~\cite{PhysRevB.44.2664,doi:10.1142/S0217979298000570}, for example through Higgs condensation in a parent U(1) theory. Both belong to the broader landscape of quantum spin liquids~\cite{Balents2010}. Related deconfined gauge theories may also emerge only near a phase transition between two symmetry-breaking phases, as in deconfined quantum critical points (DQCPs)~\cite{PhysRevB.49.11919,science.1091806,PhysRevX.5.041048}. These emergent low-energy theories can produce characteristic signatures in observables such as thermal transport and neutron-scattering spectra, which have been actively searched for experimentally~\cite{shores2005,Han2012,Zeng2024,breidenbach2025,science.adc9487}.

High-precision numerical solutions of spin-model ground states are essential for unraveling frustrated magnetism. They connect effective theories to experimental signatures in candidate systems and address the stability issue of emergent theories. Developing high-precision methods is therefore a central task in computational quantum many-body physics. While exact diagonalization (ED) is restricted to small clusters by the exponential growth of the Hilbert space, quantum Monte Carlo (QMC) can efficiently access low-temperature states as long as there is no sign problem. For the square-lattice Heisenberg model, for example, QMC can determine the ground-state energy with uncertainty at the $10^{-8}$ level~\cite{Sandvik_2026}. Frustrating interactions, however, generally introduce a sign problem, as in the $J_1$-$J_2$ model. QMC is also a major tool for lattice gauge theories, but simulations of compact QED$_3$ with massless fermions remain challenging~\cite{PhysRevD.109.034507}, leaving its infrared stability difficult to address. Variational methods offer an alternative by optimizing parametrized wavefunctions, with recent approaches employing increasingly expressive ansätze, notably tensor network states~\cite{verstraete2004,nishio2004tensor} and, more recently, neural quantum states~\cite{Chen2024MinSR,nomuraDiracTypeNodalSpin2021a, Viteritti2026ThermodynamicNQS} and their combination~\cite{Fan2026NuTNS}. Improved variational energies are generally expected to yield more accurate observables and correlations. These increasingly powerful methods are beginning to shed new light on prototypical models of frustrated quantum magnetism, racing to resolve questions that have been debated for decades.

One example is the square-lattice $J_1$-$J_2$ Heisenberg model. A long-standing debate concerns whether an intermediate $\mathbb{Z}_2$ spin liquid~\cite{PhysRevB.65.165113,PhysRevB.88.060402,PhysRevB.104.045110} exists between the Néel and valence-bond-solid (VBS) phases as $J_2$ increases. Spectroscopic analyses of the termination of Néel order~\cite{wangCriticalLevelCrossings2018,nomuraDiracTypeNodalSpin2021a,PhysRevB.109.075129,qianAbsenceSpinLiquid2024} yield estimates of the critical $J_2$ that shift substantially with increasing system size, potentially favoring a direct transition associated with a (proximate) DQCP~\cite{qianAbsenceSpinLiquid2024}. Structure-factor and order-parameter analyses~\cite{gongPlaquetteOrderedPhase2014,nomuraDiracTypeNodalSpin2021a,liuGaplessQuantumSpin2022a,Fan2026NuTNS} likewise give widely varying phase boundaries because of limited precision and strong finite-size effects. Even within the VBS regime, identifying the rotational symmetry pattern is difficult: structure-factor extrapolation diagnostics remain inconclusive~\cite{PhysRevB.74.144422}, while pinned-order studies show spurious strong local anisotropy~\cite{gongPlaquetteOrderedPhase2014,PhysRevB.110.195111}.

\begin{figure*}[t]
  \centering
  \includegraphics[width=\textwidth,trim=3bp 0bp 5.5bp 0bp,clip]{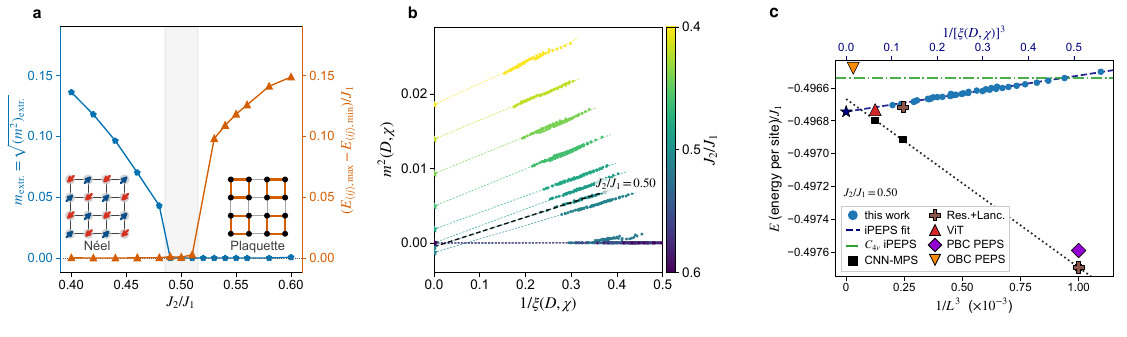}
  \caption{\textbf{\boldmath Order parameters, energies, and their extrapolation for the square lattice $J_1$-$J_2$ model.} \textbf{a} N\'eel order and plaquette order parameters as a function of $J_2/J_1$.
  Lines are guides to the eye. The dimensionless plaquette order parameter is $O_{\rm VBS}=(E_{\langle ij\rangle,\max}-E_{\langle ij\rangle,\min})/J_1$, where the maximum and minimum are taken over nearest-neighbor bond energies $E_{\langle ij\rangle}=J_1\langle \mathbf{S}_i \cdot \mathbf{S}_j \rangle$.
  In the gapped plaquette VBS phase, our data (up to $D=16$, $\chi=1024$) show convergence at the level of marker size. The corresponding bond-dimension extrapolation at $J_2/J_1=0.55$ is shown in \cref*{fig:square_j055_vbs_invD}. The N\'eel order is obtained by extrapolation.
  \textbf{b} Extrapolation of the N\'eel order from linear fits of $m^2$ versus
  $1/\xi(D,\chi)$, using optimized points with $D=8$--$16$ and $\chi=128$--$1024$.
  Stars mark the extrapolated values at $1/\xi=0$.
  \textbf{c} iPEPS extrapolation of ground-state energy per site at $J_2/J_1=0.5$ and comparison with other recent best finite-size variational results shown on the $1/L^3$ axis. Linear extrapolation is performed for $E$ versus $1/[\xi(D,\chi)]^3$.  The previous best iPEPS result~\cite{Zhang2026QRCTMRG} (horizontal line, $D=8$) is based on a $C_{4v}$ ansatz, which is a special case of our $\pi$-mirror ansatz [\cref{fig:pimirror}]. To approach the precise $E$ of the infinite system, our iPEPS extrapolation converges from above with decreasing $1/[\xi(D,\chi)]^3$. This is also the case for open boundary calculation (OBC PEPS
  \cite{Liu2021FinitePEPS}). By contrast, finite-size results with periodic boundary conditions (CNN-MPS \cite{Fan2026NuTNS}, Res.+Lanc.
  \cite{Chen2024MinSR}, ViT \cite{Viteritti2026ThermodynamicNQS}, PBC PEPS \cite{Dong2024BeyondBoundaries}) converge from below.
   The dotted line connects the lowest PBC variational energies.}
  \label{fig:fig1}
\end{figure*}

Another example is the kagome Heisenberg model. Early numerical and theoretical studies established it as a leading candidate for a quantum-disordered ground state and a U(1) Dirac spin liquid~\cite{PhysRevB.63.014413}. The theory predicts coexisting characteristic susceptibilities, including dominant oscillatory valence-bond correlations and antiferromagnetic correlations~\cite{PhysRevB.87.060405,Iqbal2015,pnas.1807840116,PhysRevLett.118.137202,Song2019}. A persistent numerical puzzle is that the predicted oscillatory valence-bond correlations are absent in several earlier~\cite{Yan2011SpinLiquidKagome,PhysRevB.77.224413} and recent~\cite{PhysRevX.15.011047} variational studies. Twisted boundary conditions have been introduced to reveal this signature, but also produce less-expected chiral correlations~\cite{he2017}. This is consistent with the strong finite-size effects indicated by ED~\cite{PhysRevB.100.155142} and theoretical analysis~\cite{PhysRevB.63.014413}. More generally, identifying a gapless spin liquid requires approaching the low-energy limit and determining whether the predicted susceptibilities coexist, or whether one of the susceptibilities instead develops into an instability toward long-range order. Their scaling dimensions provide a quantitative diagnostic.

In this work, we substantially extend the scalability of two-dimensional tensor network methods~\cite{verstraete2004,nishio2004tensor} and investigate the models above. Specifically, we use infinite projected entangled-pair states (iPEPS), which are formulated directly in the thermodynamic limit and thus give immediate access to order parameters. Their precision, however, is limited by how far the variational space, controlled by the bond dimension $D$, can be scaled up in practice. We overcome long-standing computational bottlenecks with two developments: a near-ideal parallelization scheme for modern high-performance computing architectures, and a randomized low-rank method to treat longer-range couplings more efficiently.

These advances enable high-precision calculations. In the sign-problem-free limit of the square-lattice $J_1$-$J_2$ model, we achieve precision comparable to QMC. Our results further reveal an isotropic plaquette order in the VBS regime, and we examine competing scenarios for the intermediate regime. For the kagome model, we find evidence for oscillatory valence-bond susceptibilities coexisting with other theoretically predicted susceptibilities, with extrapolations based on scaling-dimension analysis supporting a stable U(1) spin liquid.

\section{Results}
\label{sec:results}
We consider spin-$\tfrac{1}{2}$ models on square and kagome lattices with Heisenberg exchange interactions
\begin{align}
H=J_1\sum_{\langle ij \rangle}\mathbf{S}_i\cdot\mathbf{S}_j+J_2\sum_{\langle\langle ij \rangle\rangle}\mathbf{S}_i\cdot\mathbf{S}_j,
\end{align}
where $J_1>0$ is the nearest-neighbor coupling, and the next-nearest-neighbor coupling $J_2>0$ is tuned for the square lattice model to enhance frustration.

For the $J_1$-$J_2$ square lattice model, our iPEPS determination of the order parameters is presented in \cref{fig:fig1}a. The Néel antiferromagnetic order, schematically depicted in the upper-left inset, is the unfrustrated order at $J_2=0$. With increasing $J_2/J_1$, the Néel order decreases. For $J_2/J_1 \geq 0.52$, we find the emergence of plaquette valence-bond solid order, depicted in the right inset. Here, the nearest-neighbor bond energy $E_{\langle ij\rangle}=J_1\langle \mathbf{S}_i \cdot \mathbf{S}_j \rangle$ spontaneously becomes nonuniform, with each of the four low-energy bonds forming a plaquette. The pattern preserves the $C_{4v}$ rotational symmetry around the plaquette center.

\begin{figure*}[t]
  \centering

  \includegraphics[width=\textwidth]{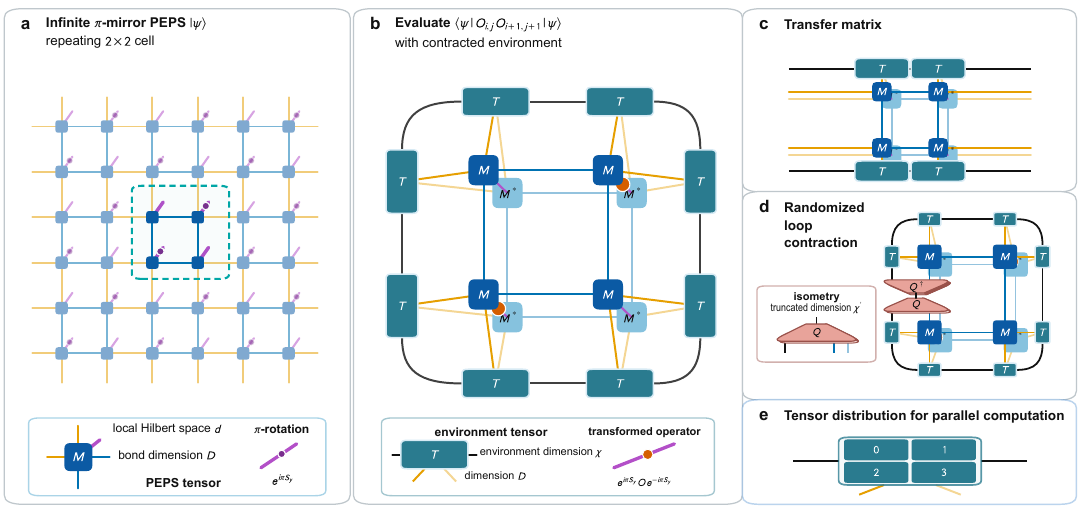}
  \caption{\textbf{Infinite variational wavefunction ansatz and tensor-network contraction for evaluating physical observables.}
\textbf{a} Infinite projected entangled pair state (iPEPS) with combined $\pi$ spin-rotation and mirror symmetries. The variational state is constructed by contracting PEPS tensors on the square lattice, with the open leg representing the local spin-$1/2$ degree of freedom. A $2\times2$ unit cell accommodates both Néel and plaquette-VBS orders. The variational parameters are encoded in a single iPEPS tensor $M$ with diagonal mirror symmetry; the remaining tensors in the unit cell are generated by mirror transformations and/or $\pi$ spin rotations.
\textbf{b} To evaluate observables supported within the unit cell, the tensors outside the cell are compressed as environment tensors $T$, obtained using the VUMPS algorithm (see Methods). The evaluation of a diagonal correlation function is illustrated; nearest-neighbor correlations and the norm are evaluated analogously.
\textbf{c} Transfer matrix of a unit cell. The correlation length $\xi$ is extracted from its leading eigenvalues.
\textbf{d},\textbf{e} Two techniques developed to improve computational scalability. \textbf{d} Randomized loop contraction reduces the cost of the direct contraction in b, with its error benchmarked against direct contraction (the construction of $Q$ is described in Methods). \textbf{e} 2DSlice partitions each environment tensor $T$ along both boundary indices to parallelize contractions and environment determination while reducing memory per device. The four numbered blocks illustrate a $2\times2$ device grid; the scheme scales well to larger grids.}
  \label{fig:pimirror}
\end{figure*}

The data are obtained using our variational iPEPS ansatz, dubbed the $\pi$-mirror ansatz and illustrated in \cref{fig:pimirror}a. The wavefunction is represented by an infinite square lattice of site tensors connected through virtual bonds. Each tensor carries one physical leg of dimension $d=2$ for the local spin-$1/2$ degree of freedom and four virtual legs of bond dimension $D$, which controls the number of variational parameters. We use a $2\times2$ unit cell, allowing for plaquette-VBS order. For technical convenience, we perform the standard sublattice $\pi$ spin rotation, under which the Néel state can be represented by a single site-centered $C_{4v}$-symmetric tensor. Our ansatz contains only one independent tensor; the other three tensors in the unit cell are generated by mirror transformations and/or the $\pi$ spin rotation. We also impose time-reversal invariance by restricting the tensor entries to real values. Besides Néel order, this ansatz accommodates plaquette, columnar, and anisotropic plaquette VBS orders. For more efficient optimization at large $D$, we further impose diagonal mirror symmetry on the independent tensor. High-precision calculations with a less-restricted ansatz show that the isotropic plaquette state is energetically favored (see Supplementary Information).

Physical observables in iPEPS can be evaluated and extrapolated with controlled precision by efficiently compressing the environment surrounding the region of interest. For example, the variational energy is obtained from expectation values $\langle \psi |\mathbf{S}_i\!\cdot\!\mathbf{S}_j| \psi \rangle/\langle \psi | \psi \rangle$ for nearest- and next-nearest-neighbor bonds. As illustrated in \cref{fig:pimirror}b, the corresponding bra-ket tensor network is contracted using environment tensors $T$, which encode the effect of tensors outside the unit-cell region. A transfer matrix, illustrated in \cref{fig:pimirror}c, is constructed by contracting a strip of iPEPS tensors with the environment in one direction while leaving virtual indices open along the others. Correlation functions are then obtained by repeated application of the transfer matrix, and a correlation length $\xi(D,\chi)$ follows from its leading eigenvalues. Ideally, $\xi(D,\chi)$ approaches the true ground-state correlation length $\xi$ as $D,\chi\rightarrow\infty$. In a VBS phase, where excitations are gapped, $\xi$ remains finite. In the Néel phase, spontaneous breaking of continuous SU(2) spin symmetry produces gapless spin-wave modes and a divergent correlation length. Likewise, gapped and gapless spin liquids are expected to exhibit finite and divergent correlation lengths, respectively. Nonetheless, at finite $D$ and environment bond dimension $\chi$, the transfer matrix has a finite effective spectrum and therefore imposes a finite correlation length $\xi(D,\chi)$. For gapped phases, $\xi(D,\chi)$ converges to the physical correlation length, as observed in our VBS data (see Supplementary Information). For gapless phases, by contrast, $\xi(D,\chi)$ acts as a soft infrared cutoff that grows with increasing $D$ and $\chi$.

This gives rise to a procedure known as finite-correlation-length scaling~\cite{PhysRevLett.129.200601,PhysRevX.8.031030,PhysRevX.8.031031}, which we use to identify gapless phases and characterize their low-energy properties. The central idea is analogous to finite-size scaling, with the soft infrared cutoff $\xi(D,\chi)$ replacing the finite system size $L$. For gapless phases with linearly dispersing excitations, including the Néel phase and Dirac spin liquids, the finite-size correction to the energy density takes the form
$E(L)=E_\infty+e_1/L^3+\cdots$~\cite{Sandvik_2026}.
Correspondingly, for iPEPS we use
$E(D,\chi)=E_\infty+e'_1/\xi(D,\chi)^3+\cdots$~\cite{PhysRevLett.129.200601,PhysRevX.8.031031}. We next consider Néel order with the staggered magnetization $\mathbf{m}_i=(-1)^{x_i+y_i}\mathbf{S}_i$. In finite systems, where spontaneous symmetry breaking cannot be directly accessed, we use the normalized structure factor $m^2(L)=N^{-2}\sum_{i,j}\langle\mathbf{m}_i\cdot\mathbf{m}_j\rangle$, with $N=L^2$, and extrapolate it as
$m^2(L)=m^2+m_1/L+\cdots$.
The leading correction follows from the long-distance magnon correlation
$\langle\mathbf{m}_i\cdot\mathbf{m}_j\rangle=m^2+a/|\mathbf{r}_i-\mathbf{r}_j|+\cdots$.
In iPEPS, by contrast, spontaneous symmetry breaking occurs directly in the thermodynamic limit, allowing the order parameter to be read out as
$m(D,\chi)=N_{\rm uc}^{-1}\sum_{\alpha=1}^{N_{\rm uc}}|\langle\mathbf{S}_\alpha\rangle_{D,\chi}|$,
where $N_{\rm uc}$ is the number of physical spin sites in the unit cell. We use this per-site amplitude for both lattices throughout the
figures. For the collinear N\'eel pattern with equal and opposite sublattice
moments, it equals $|\langle\mathbf{m}_i\rangle|$ and
$\tfrac12|\langle\mathbf{S}_A\rangle-\langle\mathbf{S}_B\rangle|$.
For kagome, the local moment directions can differ, and the average of their
magnitudes is distinct from the magnitude of their vector sum.
The corresponding finite-correlation-length scaling is
$m^2(D,\chi)=m^2+m'_1/\xi(D,\chi)+\cdots$.

For $J_2=0$, a recent QMC study~\cite{Sandvik_2026} provides an unprecedentedly precise benchmark. With our large-scale calculations, we obtain the ground-state energy at a comparable precision, reaching the $10^{-8}$ level. Further details of this benchmark are provided in Methods.

For $J_2\neq0$, we systematically improve the precision by increasing $D$ and $\chi$. \Cref{fig:fig1}b shows linear extrapolations of $m^2$ versus $1/\xi(D,\chi)$, which yield the Néel order parameter shown in \cref{fig:fig1}a. When the extrapolation instead gives an unphysical negative value of $m^2$, it indicates the absence of Néel order; this occurs in our data for $J_2/J_1\geq0.49$. At the extensively studied point $J_2/J_1=0.5$, \cref{fig:fig1}c shows the corresponding energy extrapolation. Its linear behavior over a broad range of $1/[\xi(D,\chi)]^3$ provides strong evidence for gapless excitations with linear dispersion. We also compare our results with a recent iPEPS study~\cite{Zhang2026QRCTMRG} and state-of-the-art variational calculations on finite systems~\cite{Liu2021FinitePEPS,Chen2024MinSR,Dong2024BeyondBoundaries,Viteritti2026ThermodynamicNQS,Fan2026NuTNS}. Our best finite-$D$ results already lie remarkably close to the extrapolated energies, which enables a leading resolution of the low-energy physics. A more detailed evaluation shows that the difference is below $10^{-4}$ (see Methods).

By determining the order parameters, our data identify a region [shaded in \cref{fig:fig1}a] absent of symmetry-breaking orders. Among the referenced literature, our estimate is close to the level-spectroscopy result of Ref.~\cite{nomuraDiracTypeNodalSpin2021a}, with the region notably narrower on the VBS side than in the structure-factor analysis---based on the correlation ratio---of that same reference. This smaller $J_2/J_1$ value for the VBS transition is also compatible with both earlier and more recent literature~\cite{wangCriticalLevelCrossings2018,PhysRevB.109.075129,qianAbsenceSpinLiquid2024}. Determining the boundary of the Néel phase is subject to larger uncertainty, as reflected in the large extrapolation distance (small effective correlation length) of the  Néel order in our calculation, \cref{fig:fig1}b, as well as the large size-dependent shift of boundary estimation in previous level-spectroscopy studies~\cite{wangCriticalLevelCrossings2018,nomuraDiracTypeNodalSpin2021a, PhysRevB.109.075129,qianAbsenceSpinLiquid2024}. We believe that this reflects a general numerical challenge:  the Néel order converges slowly as the low-energy limit is approached.

Beyond inferring the absence of order, unraveling spin liquids requires a direct probe of their defining characteristics. For the square-lattice model at $J_2/J_1=0.5$, our energy extrapolation analysis indicates linear gapless dispersion. This is consistent with a gapless $\mathbb{Z}_2$ spin liquid as the leading candidate for the possible intermediate symmetric phase~\cite{PhysRevB.104.045110,nomuraDiracTypeNodalSpin2021a}. Theoretically, its characteristics are coexisting Néel and VBS susceptibilities, manifesting as algebraic decay of two-point correlation functions of order parameters with characteristic exponents.  Theoretical proposals for a gapless U(1) spin liquid exist for the kagome lattice owing to the lattice obstruction to the $2\pi$ monopole instability present in the square lattice. The U(1) spin liquid is also characterized by coexisting Néel and VBS susceptibilities. Hence, before further investigation of the possible square-lattice spin liquid phase, we first present our data for the kagome model, and then comparatively investigate susceptibilities of both spin liquid candidates.

\begin{figure}[tbp]
  \centering
  \includegraphics[width=\columnwidth]{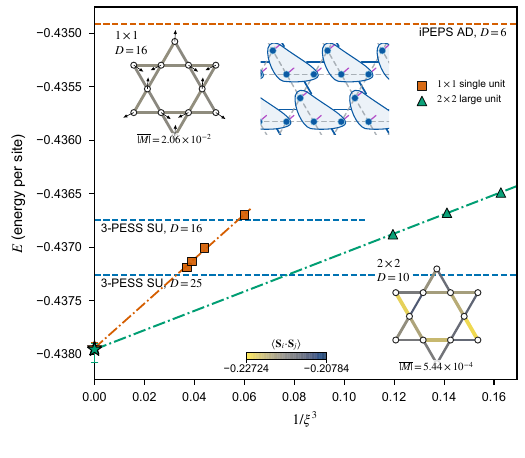}
  \caption{\textbf{iPEPS extrapolation of ground-state energy per site for the kagome-lattice Heisenberg model.} We use $1\times1$ (square marker, $D=10,12,14,16$) and $2\times2$ (triangle marker, $D=8,9,10$) unit cells. We use $\chi$ up to 1536 in the optimization. Linear extrapolation ($E$ vs. $1/[\xi(D,\chi)]^3$) is performed for both cases, and the extrapolated energies agree up to a difference $1.34\times10^{-5}$. The upper-middle inset shows the iPEPS ansatz: each tensor has a local Hilbert space of three lattice sites in a unit cell, and its four virtual bonds are contracted with the neighboring tensors; dashed lines denote the kagome lattice defining physical neighbors for couplings. The $1\times1$ unit cell is compatible with  120$\degree$ AFM susceptibilities (upper-left inset);  $2\times2$ unit cell state exhibits VBS susceptibilities (lower-right inset). These susceptibilities manifest as finite-$D$ order parameters that tend to vanish with increasing $\xi(D,\chi)$ ($D$); see \cref{fig:square_kagome_m2_xi}. Horizontal dashed lines show the previous gradient-optimized $D=6$ iPEPS result~\cite{hu2026dynamicalspectralfunctionkagome} and the infinite projected entangled-simplex state (3-iPESS) results optimized by approximate imaginary time evolution at $D=16$ and $25$~\cite{PhysRevLett.118.137202}.
  }
  \label{fig:kagome_energy_xi3}
\end{figure}

For the kagome lattice model, our data are based on optimizing a sublattice-merged iPEPS~\cite{PhysRevB.86.041106} (\cref{fig:kagome_energy_xi3}, upper-middle inset). We consider both a fully translation-invariant ansatz and a two-by-two enlarged unit cell. As expected, the 120$\degree$  AFM order (\cref{fig:kagome_energy_xi3}, upper-left inset) emerges in the fully translation-invariant ansatz, consistent with previous infinite tensor network studies~\cite{PhysRevLett.118.137202, hu2026dynamicalspectralfunctionkagome}. Finite-$D$ order that vanishes with increasing  $D$ is an indication of corresponding susceptibilities. Our results for the two-by-two enlarged unit cell exhibit a much smaller magnetization and 
finite-$D$ valence-bond solid order (\cref{fig:kagome_energy_xi3}, lower-right inset). This is consistent with the oscillatory valence-bond susceptibilities predicted for U(1) spin liquids. Absence of evidence of the susceptibilities in previous numerical studies without twisted boundary conditions might be a finite-size effect. The VBS susceptibility is related to the gapless excitation at M points of the Brillouin zone found in tensor network studies~\cite{pnas.1807840116,hu2026dynamicalspectralfunctionkagome}. The energy extrapolation in $1/[\xi(D,\chi)]^3$ is shown in \cref{fig:kagome_energy_xi3}, consistent with gapless linear dispersion. Notably, the two different choices of unit cell yield nearly identical extrapolated results. We plot previous tensor network results for comparison.

\begin{figure}[tbp]
  \centering
  \includegraphics[width=\columnwidth]{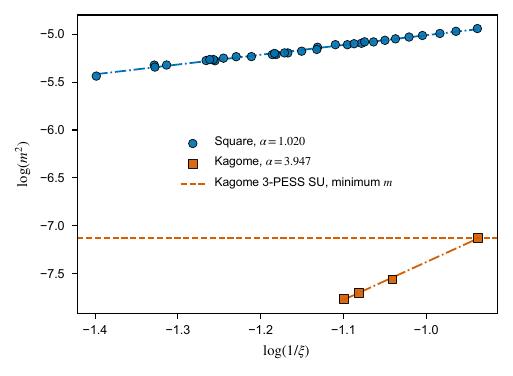}
  \caption{\textbf{\boldmath Inference of scaling dimension of Néel operators for the square-lattice
  $J_1$-$J_2$ model at $J_2/J_1=0.5$ and the kagome Heisenberg
  model with $1\times1$ unit cell.} Here $m$ is the average local spin-moment
  magnitude per physical site, and $m^2$ is the square of that average.
  Linear fits to $\log(m^2)$ versus $\log(1/\xi)$ give slopes ($\alpha$)
  $1.020(16)$ and $3.95(11)$, an estimate for $2\Delta_m$.
  The horizontal dashed line marks the lowest magnetization in a
  previous iPESS result~\cite{PhysRevLett.118.137202}.
  }
  \label{fig:square_kagome_m2_xi}
\end{figure}

We can now analyze the susceptibilities of the gapless spin liquid candidates, probing the characteristic signatures of the gapless phases. In \cref{fig:square_kagome_m2_xi},
we plot $m^2$ vs. $1/\xi$ on a log-log scale for the $J_1$-$J_2$ square lattice Heisenberg model with $J_2/J_1=0.5$ and the kagome lattice Heisenberg model with $1\times 1$ unit cell. Assuming gapless spin liquids, we thus use finite-correlation-length scaling~\cite{PhysRevX.8.031030,PhysRevX.8.031031,PhysRevB.104.195126}, which requires the well-converged variational iPEPS data that we have access to. We fit
\begin{align}\label{fcs}
m^2(D,\chi) \propto [1/\xi(D,\chi)]^{2\Delta_m},
\end{align}
where $\Delta_m$ is the dominant scaling dimension of the spin operator. The fits yield $2\Delta_m = 1.020(16)$ for the square lattice and $2\Delta_m = 3.95(11)$ for the kagome lattice. Using $2\Delta_m$
as a proxy for the infrared scaling dimension of the spin–spin coupling term, the kagome value exceeds the space-time dimension $d+1=3$, so this term is irrelevant, consistent with a stable gapless phase. On the square lattice, by contrast, $2\Delta_m < 3$, so the coupling is relevant and destabilizes the gapless state. For comparison, previous estimates for the square lattice, from fits to real-space correlation functions~\cite{nomuraDiracTypeNodalSpin2021a,liuGaplessQuantumSpin2022a} or from finite-size structure factors~\cite{Fan2026NuTNS}, all point to $2\Delta_m<1.5$. Assuming an intermediate spin-liquid phase, we attribute these small values to the proximity of the Néel–spin-liquid critical point, or even the Néel phase itself, which may still dominate the correlations on the energy scales accessible to numerics.
Effective field theory provides further insight.
For both $\mathbb{Z}_2$ and U(1) spin liquids, the bilinears of emergent fermions contribute to the spin-spin correlation functions.  For gapless $\mathbb{Z}_2$ spin-liquid phases, scaling dimensions of bilinears are expected to follow those of free Dirac fermions, predicting $\langle \mathbf{m}_i \cdot \mathbf{m}_j\rangle=a/|\mathbf{r}_i-\mathbf{r}_j|^4+\cdots$, namely $2\Delta_m=4$. However, for a U(1) spin liquid, the bilinears may have different scaling dimensions, and there is also a possibly more dominant $4\pi$-flux monopole contribution~\cite{Song2019}. Its scaling dimension $\Delta_M$ is not precisely determined. 
The large-$N_f$ expansions of compact QED$_3$, uncontrolled at $N_f=4$ flavors, yield $2\Delta_M \approx 2.5$~\cite{Song2019}, which is relevant and could destabilize QED$_3$.
Nevertheless, the proxy is close to the threshold 3. On the other hand, QMC~\cite{PhysRevD.109.034507} can only efficiently simulate the noncompact theory, giving $2\Delta_M \approx3.7(3)$ as another proxy. Since our fitting result for the dominant $2\Delta_m$ of the kagome model is close to 4, it indicates that the monopole effects are either subdominant (irrelevant) or that the term has a small prefactor. In the Supplementary Information, we present our $2\times2$-unit-cell data for VBS susceptibilities.

In summary, by improving the scalability of the iPEPS method, we obtain high-precision data, which provide insights through extrapolation, including inferring the scaling dimension of susceptibilities.  Our analysis indicates a possibly stable gapless spin liquid on the kagome lattice, and competing scenarios of the square lattice model near $J_2/J_1=0.5$. The gapless spin liquids are academically defined through the low-energy (infrared) limit. On the other hand, there are infrared cutoffs in the real world such as temperature and neutron scattering resolution. By approaching similar energy resolution, the iPEPS method can serve as a guide for spin liquid identification.

\section{Methods}
\label{sec:method}

We implement and develop numerical methods to evaluate and optimize two-dimensional tensor networks. The ingredients are the tensor network ansatz itself, the environment tensor determination, the evaluation of energies and other observables, the optimization, and the extrapolation of the results.
As we show in \cref{fig:pimirror}b, observable evaluations use environments as input. Once the energy is evaluated,
automatic differentiation (AD)~\cite{ADPEPS} can generate energy gradients for energy-minimization optimization~\cite{VPEPS,ADPEPS,PhysRevResearch.7.013237,burgelman2026implicitdifferentiationtensornetwork}, for which we use L-BFGS with a local preconditioner~\cite{h396-yc28}.  

The iPEPS tensor fully encodes the variational wavefunction. The variational parameters correspond to the tensor entries, the number of which is proportional to $d D^4$. The variational manifold is enlarged with increasing $D$ and thereby enables better-optimized variational energies.  The number of parameters of the environment tensors, proportional to $\chi^2 D^2$,  controls the precision of the energies, order parameters, and correlation lengths of an iPEPS. We label the calculated values with the corresponding $D$ and $\chi$.  Exact evaluations of iPEPS expectation values are ideally obtained in the limit $\chi_{\mathrm{ev}} \rightarrow \infty$, e.g., for the energy $E(D,\chi_{\mathrm{opt}}, \chi_{\mathrm{ev}}=\infty)$. Note that here, we allow for using a different $\chi_{\mathrm{opt}}$ to compute the energy and its gradient during the optimization process, and could choose to only extrapolate to large $\chi_{\mathrm{ev}}$ at convergence. A second extrapolation $D,\chi_{\mathrm{opt}} \rightarrow\infty$ is needed.
Here, we do not employ this approach, and rather use a one-shot extrapolation scheme~\cite{PhysRevLett.129.200601} that takes data optimized and evaluated with the same $\chi$. No accompanying $\chi_{\mathrm{ev}}$ extrapolation is needed. The example in \cref*{fig:chi_reoptimization} shows how these two schemes can induce different extrapolation results. For gapless phases, the one-shot extrapolation scheme assumes that the finite $D$ and $\chi$ as a deviation from the exact limit can be unified to the soft IR cutoff $\xi(D,\chi)$, provided we indeed use $\chi_{\mathrm{opt}} = \chi_{\mathrm{ev}}=\chi$.  More specifically, our data for the $J_1$-$J_2$ model indicate that $E(D,\chi)$ is approaching the exact values monotonically from above in both $\chi$ and $D$, for sufficiently large $\chi$. This is indeed compatible with the extrapolation scheme and the observation that $\xi(D,\chi)$ grows with $D$ and $\chi$. The monotonicity has been observed previously in studies based on $C_{4v}$ iPEPS~\cite{ADPEPS,PhysRevLett.129.200601} and iPEPS with mirror symmetry~\cite{PhysRevB.105.195140}, which contains our $\pi$-mirror ansatz. The mirror symmetry in these ansätze makes the norm converge monotonically from below in $\chi$~\cite{PhysRevB.105.195140}, and is therefore likely relevant to the energy convergence behavior in $\chi$ as well. Since finite-$\chi$ evaluation practically induces energy overestimation for $\pi$-mirror and $C_{4v}$ iPEPS, we can get a better estimate of the energy for a given iPEPS by scaling up $\chi$. For example, for $J_2/J_1=0.5$, the iPEPS has been optimized and evaluated at $D=16,\chi=1024$; using $\chi=2048$ for evaluation, we obtain $E=-0.4967113$, closer to the extrapolated result $E=-0.4967446(18)$. Nonetheless, as mentioned above, the scheme we adopt uses data evaluated for the same value of $\chi$ used during the optimization, but does allow us to collect data with different $\chi_{\mathrm{opt}} = \chi_{\mathrm{ev}}$ for a given value of $D$. The quality of the linear extrapolation of our $J_1$-$J_2$ data (\cref{fig:fig1}) demonstrates the self-consistency of this one-shot unified scheme.
Note that the mirror symmetry is absent in our iPEPS ansatz for the kagome model, and it turns out that $E(D,\chi_{\mathrm{opt}})$ may instead be smaller than $E(D,\chi_{\mathrm{opt}}, \chi_{\mathrm{ev}}=\infty)$. Therefore, we use sufficiently large environment bond dimensions during optimization and employ the minimal canonical form~\cite{Acuaviva2023MinimalCanonicalForm,Tang2025Gauging} to improve numerical conditioning, rendering $\chi$-deviation effects negligible. Only one $\chi$ is used per $D$ in \cref{fig:kagome_energy_xi3}.

\begin{figure}[tbp]
  \centering
  \includegraphics[width=\columnwidth]{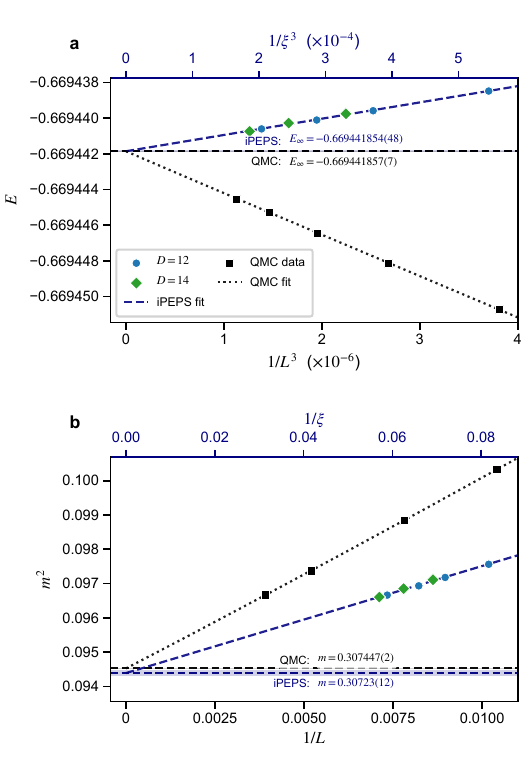}
  \caption{\textbf{Extrapolations of the ground-state energy and N\'eel order of the square-lattice Heisenberg model, comparing our $C_{4v}$ iPEPS calculations with the QMC reference~\cite{Sandvik_2026}.} We use $D=12,14$ and $\chi$ between $512$ and $1536$. For iPEPS, $m^2$ is the square of the per-site N\'eel amplitude $m=\tfrac12|\langle\mathbf{S}_A\rangle-\langle\mathbf{S}_B\rangle|$. For QMC, $m^2(L)=\langle\bar{\mathbf m}\cdot\bar{\mathbf m}\rangle$, with $\bar{\mathbf m}=L^{-2}\sum_i(-1)^{x_i+y_i}\mathbf S_i$, on periodic $L\times L$ lattices at zero or sufficiently low temperature. The QMC reference extends to $L=96$ for energy and $L=256$ for $m^2$. The iPEPS fits use only the leading corrections in $1/\xi^3$ and $1/\xi$, respectively; the QMC fits include the subleading corrections reported in the reference.}
  \label{fig:squareheisenberg}
\end{figure}

As another demonstration and benchmark, we present in \cref{fig:squareheisenberg} our $C_{4v}$ iPEPS result for the Heisenberg model ($J_2=0$) in comparison with the recent high-precision QMC reference~\cite {Sandvik_2026}. Similar to \cref{fig:fig1}, we use the system width ($L$) of the finite-size QMC results and the soft IR cutoff [$\xi(D,\chi)$] of our iPEPS calculations as length scales to organize the plots. Note that in \cref{fig:fig1}, variational finite-size results from the literature were not extrapolated to $L \rightarrow \infty$ because the variational errors were not guaranteed to be sufficiently small. In contrast, for the QMC results for the Heisenberg model, the sampled energies have error bars as small as $10^{-7}$, and extrapolations with subleading corrections are performed in Ref.~\cite{Sandvik_2026} and replicated in \cref{fig:squareheisenberg}.
In order to extrapolate our iPEPS results, we control their quality by insisting that the energy gradients for this model are as small as $10^{-6}$. In the plot, we use blue and green to denote $D=12$ and $D=14$ data, optimized and evaluated with a series of $\chi$. The energies decrease monotonically with both $D$ and $\chi$.  Note that we used even larger sets (and values) of $D$ and $\chi$ for \cref{fig:fig1}. The complete listing of the results per $(D, \chi)$ pair can be found in the raw data set.  The extrapolation only uses the leading term, i.e.\ $1/\xi(D,\chi)^3$ for energy and $1/\xi(D,\chi)$ for magnetization. The extrapolated energies agree with QMC up to 8 significant digits. Furthermore, our largest $(D,\chi)$ result lies closer to the extrapolated energy than the result of the largest QMC system size. The precision is significantly improved compared to the previous implementation of the same extrapolation scheme~\cite{PhysRevLett.129.200601}. This demonstrates the improved iPEPS data quality that is enabled by working with large $D$ and $\chi$.

However, this up-scaling of $D$ and $\chi$ requires overcoming some key challenges, namely reducing the computational cost and being able to effectively utilize modern parallel computation architectures. These are known as bottlenecks of iPEPS methods. To achieve the latter, we introduce a scheme that realizes nearly ideal parallelization of the evaluation and optimization. To further help with the computational cost, we also propose a randomized loop contraction scheme for evaluating the next-nearest-neighbor coupling in the $J_1$-$J_2$ model, which reduces complexity while maintaining sufficient precision.

We determine the environment tensor using variational uniform matrix product states (VUMPS) algorithms~\cite{PhysRevB.97.045145,Nietner2020efficient,PhysRevB.108.085103,Zhang2025HigherSpinKitaev}. Compared with the more conventional corner transfer matrix renormalization group (CTMRG) algorithm~\cite{CTM,PhysRevLett.113.046402}, VUMPS does not require large-scale projector construction, and its computational cost is tensor-contraction dominated, with subleading costs associated with QR decompositions. The cost complexity of each iteration step is lower than for conventional CTMRG schemes, and scales as $O(D^4\chi^3)+O(D^6\chi^2d)$~\cite{Zhang2025HigherSpinKitaev}. With the empirical choice $\chi= O(D^2)$, the cost scales as $O(D^{10}) \approx O(N^{2.5})$, where $N$ is the number of variational parameters. Compared with CTMRG with revised projector solvers~\cite{PhysRevB.107.054424,richards2025acetngpuacceleratedcornertransfermatrixrenormalization,PhysRevB.111.235116,zhang2026fasttwodimensionaltensornetworkcontraction}, VUMPS can remain faster in practice~\footnote{We limit our discussion to solving the conventional bilayer environment rather than the proxy layer-truncated environment, e.g.,~\cite{stagger}.}

\begin{figure}[H]
  \centering
  \includegraphics[width=0.62\columnwidth]{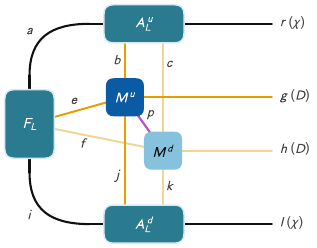}
  \caption{\textbf{\boldmath Tensor diagram of the left-environment map
  $F_L\mapsto F'_L$.} Black, orange, and purple bonds carry dimensions
  $\chi$, $D$, and $d$, respectively. The square tensors $M^u$ and $M^d$
  denote the upper and lower iPEPS layers. Repeated labels denote contracted
  indices; the four dimension-labeled open legs form the output tensor $F'_L$. Starting with $A^d_L$, the tensors $F_L$, $M^u$, $M^d$, and $A_L^u$ are contracted in this order.}
  \label{fig:flmap_contraction}
\end{figure}

The principle behind the determination of the environment is to maximize the norm (per site) of iPEPS $|\langle \psi |\psi\rangle|$. In VUMPS, this amounts to sequentially finding the dominant eigenvector, for which the elementary step is to perform a linear map to an environment tensor. We first discuss the implementation of this step with a simple tensor parallelization scheme, which we dub 1DSlice. The linear map for the left environment (denoted as $F_L$) is illustrated in \cref{fig:flmap_contraction}.
Contracting $F_L$ with the ket and bra iPEPS tensors and the top and bottom environment tensors gives the updated left environment tensor $F'_L$. The optimal order is denoted in \cref{fig:flmap_contraction}. Note that the rank-four tensor ${F'_{L}}$ can be obtained by computing the rank-three tensor slices ${F'_{L,l=i}}$ for each $i=1,..,\chi$ independently. Computing these components one by one, instead of simultaneously, has the benefit of reducing the peak memory from $O(D^4\chi^2)$ to $O(D^2\chi^2)$. Such a common multi-tensor contraction trick has been known to the tensor network community~\cite{PhysRevB.86.045139}. We find that the independence of these tasks is naturally suitable for parallel computing, with a communication cost of  $O(D^2\chi^2)$ for the full process, since each computation device should have the full updated $F_L$ in order to start the next iteration. Because the computational cost scales as $O(D^4\chi^3)+O(D^6\chi^2d)$, the relative communication cost shrinks quickly when scaling up $D$ and $\chi$. Similar linear maps appear in other tensor network algorithms such as DMRG and CTMRG, for which tensor parallelization schemes exist for solving MPS (e.g.,~\cite{Menczer2025}) and 2D classical models~\cite{PhysRevE.102.032130}. What makes our scheme fairly simple and effective for iPEPS is the $D$ scaling specific to iPEPS, and VUMPS being dominated by the application of this linear map. In this 1DSlice scheme, memory is not distributed, in the sense that each computation device keeps the same data as single-device computation. In our implementation, this approach is sufficient for calculations reaching up to $D=12$ using 64GB device memory. To scale up $D$ further and lift the bottleneck from single-device memory, we furthermore introduce a distributed memory scheme, which we dub 2DSlice (see Supplementary Information). These two schemes can be naturally extended to tensors with internal symmetries.

After determining the environments, the calculation of energies is performed term by term, each of which corresponds to a tensor contraction. With translation invariance, we sum over all couplings involving sites in one unit cell. For example, to compute the next-nearest neighbor (diagonal) coupling, we consider a cell of two-by-two bra-ket structure of iPEPS tensors contracted with surrounding environment tensors and with operators inserted in a diagonal pair of iPEPS physical legs, as was illustrated in \cref{fig:pimirror}b. Similarly, to compute the nearest-neighbor coupling, environments enclose two neighboring sites. The cost for computing the contraction for nearest-neighbor coupling scales as $O(D^4\chi^3)+O(D^6\chi^2d)$, which is the same complexity as the VUMPS iteration. However, direct contraction for the next-nearest neighbor coupling incurs higher complexity $O(D^6\chi^3)$. We introduce a randomized loop contraction scheme to reduce the cost to $O(D^4\chi^3)$. This method finds a projector $Q$, shown in \cref{fig:pimirror}d, to approximate the contraction while keeping high precision.  For each corner of \cref{fig:pimirror}b, a bra-ket pair of iPEPS tensors contracts with two environment tensors, resulting in a corner tensor with dimension $[D,D,\chi,D,D,\chi]$. Grouping the legs, they can be viewed as a $D^2\chi$ by $D^2\chi$ matrix  $\mathcal M$ and $\mathcal M_o$, where $o$ represents the inserted operator. The evaluation of \cref{fig:pimirror}b can be regarded as a loop contraction $\operatorname{Tr}(\mathcal M\mathcal M_{o}\mathcal M\mathcal M_{o})$. We replace this expression by $\operatorname{Tr}(QQ^{\dagger}\mathcal M\mathcal M_{o}\mathcal M\mathcal M_{o})$, where $Q$ is the isometric tensor in \cref{fig:pimirror}d, viewed as a $D^2\chi$ by $\bar{\chi}$ matrix. Here, $\bar{\chi}$ is the control parameter that affects the accuracy of the approximation. Using the randomized QB method \cite{Halko2011Randomized}, $Q$ can be obtained from the QR decomposition $\mathcal M\mathcal M_{o}\mathcal M\mathcal M_{o}\Omega=QR$, where $\Omega$ is a $D^2\chi$ by $\bar{\chi}$ matrix with random Gaussian entries. For the $J_1$-$J_2$ model, we find that $\bar{\chi}=O(\chi)$ is sufficient to maintain the precision $10^{-8}$. The projector construction and its numerical validation are detailed in the Supplementary Information. This reveals a hidden approximate low-rank property, which may be general beyond this model.

\paragraph{Acknowledgments.---}
We thank Nick Bultinck, Lander Burgelman, Stefano Carignano, Zheng-Cheng Gu, Naichao Hu, Yasir Iqbal, Frank Verstraete, Zhi-Yuan Xie, and Qi Yang for discussion.  We thank Hai-Jun Liao for providing the original iPESS data used for comparison in \cref{fig:kagome_energy_xi3,fig:square_kagome_m2_xi}.
The open-source Julia package TeneT.jl is available on GitHub~\cite{TeneTjl}. The data underlying all figures and tables, together with the scripts that produce them, are available on GitHub~\cite{ScalableTNData}. We used AI tools to help optimize the CUDA-NCCL deployment and the overlap of communication and computation in our parallelization schemes on HPC systems.
We acknowledge the EuroHPC Joint Undertaking for awarding projects EHPC-DEV-2025D07-003 and EHPC-DEV-2026D04-006 access to MareNostrum~5, hosted by the Barcelona Supercomputing Center (BSC), Spain, and project EHPC-DEV-2026D01-011 access to JUPITER, hosted by the J\"ulich Supercomputing Centre (JSC) at Forschungszentrum J\"ulich, Germany. We acknowledge computing resources and services provided by the VSC (Flemish Supercomputer Center), in particular, access to the full Sofia cluster for development and benchmarking. This work was supported by the European Union through the European Research Council (ERC, GaMaTeN, Grant No. 101125822). Views and opinions expressed are, however, those of the authors only and do not necessarily reflect those of the European Union or the European Research Council Executive Agency. Neither the European Union nor the granting authority can be held responsible for them.

\bibliography{references}

\begin{thebibliography}{76}%
\makeatletter
\providecommand \@ifxundefined [1]{%
 \@ifx{#1\undefined}
}%
\providecommand \@ifnum [1]{%
 \ifnum #1\expandafter \@firstoftwo
 \else \expandafter \@secondoftwo
 \fi
}%
\providecommand \@ifx [1]{%
 \ifx #1\expandafter \@firstoftwo
 \else \expandafter \@secondoftwo
 \fi
}%
\providecommand \natexlab [1]{#1}%
\providecommand \enquote  [1]{``#1''}%
\providecommand \bibnamefont  [1]{#1}%
\providecommand \bibfnamefont [1]{#1}%
\providecommand \citenamefont [1]{#1}%
\providecommand \href@noop [0]{\@secondoftwo}%
\providecommand \href [0]{\begingroup \@sanitize@url \@href}%
\providecommand \@href[1]{\@@startlink{#1}\@@href}%
\providecommand \@@href[1]{\endgroup#1\@@endlink}%
\providecommand \@sanitize@url [0]{\catcode `\\12\catcode `\$12\catcode
  `\&12\catcode `\#12\catcode `\^12\catcode `\_12\catcode `\%12\relax}%
\providecommand \@@startlink[1]{}%
\providecommand \@@endlink[0]{}%
\providecommand \url  [0]{\begingroup\@sanitize@url \@url }%
\providecommand \@url [1]{\endgroup\@href {#1}{\urlprefix }}%
\providecommand \urlprefix  [0]{URL }%
\providecommand \Eprint [0]{\href }%
\providecommand \doibase [0]{https://doi.org/}%
\providecommand \selectlanguage [0]{\@gobble}%
\providecommand \bibinfo  [0]{\@secondoftwo}%
\providecommand \bibfield  [0]{\@secondoftwo}%
\providecommand \translation [1]{[#1]}%
\providecommand \BibitemOpen [0]{}%
\providecommand \bibitemStop [0]{}%
\providecommand \bibitemNoStop [0]{.\EOS\space}%
\providecommand \EOS [0]{\spacefactor3000\relax}%
\providecommand \BibitemShut  [1]{\csname bibitem#1\endcsname}%
\let\auto@bib@innerbib\@empty
\bibitem [{\citenamefont {Baskaran}\ and\ \citenamefont
  {Anderson}(1988)}]{PhysRevB.37.580}%
  \BibitemOpen
  \bibfield  {author} {\bibinfo {author} {\bibfnamefont {G.}~\bibnamefont
  {Baskaran}}\ and\ \bibinfo {author} {\bibfnamefont {P.~W.}\ \bibnamefont
  {Anderson}},\ }\bibfield  {title} {\bibinfo {title} {{Gauge theory of
  high-temperature superconductors and strongly correlated Fermi systems}},\
  }\href {https://doi.org/10.1103/PhysRevB.37.580} {\bibfield  {journal}
  {\bibinfo  {journal} {Phys. Rev. B}\ }\textbf {\bibinfo {volume} {37}},\
  \bibinfo {pages} {580(R)} (\bibinfo {year} {1988})}\BibitemShut {NoStop}%
\bibitem [{\citenamefont {Wen}(1991)}]{PhysRevB.44.2664}%
  \BibitemOpen
  \bibfield  {author} {\bibinfo {author} {\bibfnamefont {X.~G.}\ \bibnamefont
  {Wen}},\ }\bibfield  {title} {\bibinfo {title} {{Mean-field theory of
  spin-liquid states with finite energy gap and topological orders}},\ }\href
  {https://doi.org/10.1103/PhysRevB.44.2664} {\bibfield  {journal} {\bibinfo
  {journal} {Phys. Rev. B}\ }\textbf {\bibinfo {volume} {44}},\ \bibinfo
  {pages} {2664} (\bibinfo {year} {1991})}\BibitemShut {NoStop}%
\bibitem [{\citenamefont {Balents}\ \emph {et~al.}(1998)\citenamefont
  {Balents}, \citenamefont {Fisher},\ and\ \citenamefont
  {Nayak}}]{doi:10.1142/S0217979298000570}%
  \BibitemOpen
  \bibfield  {author} {\bibinfo {author} {\bibfnamefont {L.}~\bibnamefont
  {Balents}}, \bibinfo {author} {\bibfnamefont {M.~P.~A.}\ \bibnamefont
  {Fisher}},\ and\ \bibinfo {author} {\bibfnamefont {C.}~\bibnamefont
  {Nayak}},\ }\bibfield  {title} {\bibinfo {title} {Nodal liquid theory of the
  pseudo-gap phase of high-{$T_c$} superconductors},\ }\href
  {https://doi.org/10.1142/S0217979298000570} {\bibfield  {journal} {\bibinfo
  {journal} {International Journal of Modern Physics B}\ }\textbf {\bibinfo
  {volume} {12}},\ \bibinfo {pages} {1033} (\bibinfo {year}
  {1998})}\BibitemShut {NoStop}%
\bibitem [{\citenamefont {Balents}(2010)}]{Balents2010}%
  \BibitemOpen
  \bibfield  {author} {\bibinfo {author} {\bibfnamefont {L.}~\bibnamefont
  {Balents}},\ }\bibfield  {title} {\bibinfo {title} {Spin liquids in
  frustrated magnets},\ }\href {https://doi.org/10.1038/nature08917} {\bibfield
   {journal} {\bibinfo  {journal} {Nature}\ }\textbf {\bibinfo {volume}
  {464}},\ \bibinfo {pages} {199} (\bibinfo {year} {2010})}\BibitemShut
  {NoStop}%
\bibitem [{\citenamefont {Chubukov}\ \emph {et~al.}(1994)\citenamefont
  {Chubukov}, \citenamefont {Sachdev},\ and\ \citenamefont
  {Ye}}]{PhysRevB.49.11919}%
  \BibitemOpen
  \bibfield  {author} {\bibinfo {author} {\bibfnamefont {A.~V.}\ \bibnamefont
  {Chubukov}}, \bibinfo {author} {\bibfnamefont {S.}~\bibnamefont {Sachdev}},\
  and\ \bibinfo {author} {\bibfnamefont {J.}~\bibnamefont {Ye}},\ }\bibfield
  {title} {\bibinfo {title} {{Theory of two-dimensional quantum Heisenberg
  antiferromagnets with a nearly critical ground state}},\ }\href
  {https://doi.org/10.1103/PhysRevB.49.11919} {\bibfield  {journal} {\bibinfo
  {journal} {Phys. Rev. B}\ }\textbf {\bibinfo {volume} {49}},\ \bibinfo
  {pages} {11919} (\bibinfo {year} {1994})}\BibitemShut {NoStop}%
\bibitem [{\citenamefont {Senthil}\ \emph {et~al.}(2004)\citenamefont
  {Senthil}, \citenamefont {Vishwanath}, \citenamefont {Balents}, \citenamefont
  {Sachdev},\ and\ \citenamefont {Fisher}}]{science.1091806}%
  \BibitemOpen
  \bibfield  {author} {\bibinfo {author} {\bibfnamefont {T.}~\bibnamefont
  {Senthil}}, \bibinfo {author} {\bibfnamefont {A.}~\bibnamefont {Vishwanath}},
  \bibinfo {author} {\bibfnamefont {L.}~\bibnamefont {Balents}}, \bibinfo
  {author} {\bibfnamefont {S.}~\bibnamefont {Sachdev}},\ and\ \bibinfo {author}
  {\bibfnamefont {M.~P.~A.}\ \bibnamefont {Fisher}},\ }\bibfield  {title}
  {\bibinfo {title} {{Deconfined Quantum Critical Points}},\ }\href
  {https://doi.org/10.1126/science.1091806} {\bibfield  {journal} {\bibinfo
  {journal} {Science}\ }\textbf {\bibinfo {volume} {303}},\ \bibinfo {pages}
  {1490} (\bibinfo {year} {2004})}\BibitemShut {NoStop}%
\bibitem [{\citenamefont {Nahum}\ \emph {et~al.}(2015)\citenamefont {Nahum},
  \citenamefont {Chalker}, \citenamefont {Serna}, \citenamefont {Ortu{\~n}o},\
  and\ \citenamefont {Somoza}}]{PhysRevX.5.041048}%
  \BibitemOpen
  \bibfield  {author} {\bibinfo {author} {\bibfnamefont {A.}~\bibnamefont
  {Nahum}}, \bibinfo {author} {\bibfnamefont {J.~T.}\ \bibnamefont {Chalker}},
  \bibinfo {author} {\bibfnamefont {P.}~\bibnamefont {Serna}}, \bibinfo
  {author} {\bibfnamefont {M.}~\bibnamefont {Ortu{\~n}o}},\ and\ \bibinfo
  {author} {\bibfnamefont {A.~M.}\ \bibnamefont {Somoza}},\ }\bibfield  {title}
  {\bibinfo {title} {{Deconfined Quantum Criticality, Scaling Violations, and
  Classical Loop Models}},\ }\href {https://doi.org/10.1103/PhysRevX.5.041048}
  {\bibfield  {journal} {\bibinfo  {journal} {Phys. Rev. X}\ }\textbf {\bibinfo
  {volume} {5}},\ \bibinfo {pages} {041048} (\bibinfo {year}
  {2015})}\BibitemShut {NoStop}%
\bibitem [{\citenamefont {Shores}\ \emph {et~al.}(2005)\citenamefont {Shores},
  \citenamefont {Nytko}, \citenamefont {Bartlett},\ and\ \citenamefont
  {Nocera}}]{shores2005}%
  \BibitemOpen
  \bibfield  {author} {\bibinfo {author} {\bibfnamefont {M.~P.}\ \bibnamefont
  {Shores}}, \bibinfo {author} {\bibfnamefont {E.~A.}\ \bibnamefont {Nytko}},
  \bibinfo {author} {\bibfnamefont {B.~M.}\ \bibnamefont {Bartlett}},\ and\
  \bibinfo {author} {\bibfnamefont {D.~G.}\ \bibnamefont {Nocera}},\ }\bibfield
   {title} {\bibinfo {title} {{A Structurally Perfect $S=1/2$ {Kagom\'e}
  Antiferromagnet}},\ }\href {https://doi.org/10.1021/ja053891p} {\bibfield
  {journal} {\bibinfo  {journal} {J. Am. Chem. Soc.}\ }\textbf {\bibinfo
  {volume} {127}},\ \bibinfo {pages} {13462} (\bibinfo {year}
  {2005})}\BibitemShut {NoStop}%
\bibitem [{\citenamefont {Han}\ \emph {et~al.}(2012)\citenamefont {Han},
  \citenamefont {Helton}, \citenamefont {Chu}, \citenamefont {Nocera},
  \citenamefont {Rodriguez-Rivera}, \citenamefont {Broholm},\ and\
  \citenamefont {Lee}}]{Han2012}%
  \BibitemOpen
  \bibfield  {author} {\bibinfo {author} {\bibfnamefont {T.-H.}\ \bibnamefont
  {Han}}, \bibinfo {author} {\bibfnamefont {J.~S.}\ \bibnamefont {Helton}},
  \bibinfo {author} {\bibfnamefont {S.}~\bibnamefont {Chu}}, \bibinfo {author}
  {\bibfnamefont {D.~G.}\ \bibnamefont {Nocera}}, \bibinfo {author}
  {\bibfnamefont {J.~A.}\ \bibnamefont {Rodriguez-Rivera}}, \bibinfo {author}
  {\bibfnamefont {C.}~\bibnamefont {Broholm}},\ and\ \bibinfo {author}
  {\bibfnamefont {Y.~S.}\ \bibnamefont {Lee}},\ }\bibfield  {title} {\bibinfo
  {title} {{Fractionalized excitations in the spin-liquid state of a
  kagome-lattice antiferromagnet}},\ }\href
  {https://doi.org/10.1038/nature11659} {\bibfield  {journal} {\bibinfo
  {journal} {Nature}\ }\textbf {\bibinfo {volume} {492}},\ \bibinfo {pages}
  {406} (\bibinfo {year} {2012})}\BibitemShut {NoStop}%
\bibitem [{\citenamefont {Zeng}\ \emph {et~al.}(2024)\citenamefont {Zeng},
  \citenamefont {Zhou}, \citenamefont {Zhou}, \citenamefont {Han},
  \citenamefont {Chi}, \citenamefont {Li}, \citenamefont {Kofu}, \citenamefont
  {Nakajima}, \citenamefont {Wei}, \citenamefont {Zhang}, \citenamefont
  {Mazzone}, \citenamefont {Meng},\ and\ \citenamefont {Li}}]{Zeng2024}%
  \BibitemOpen
  \bibfield  {author} {\bibinfo {author} {\bibfnamefont {Z.}~\bibnamefont
  {Zeng}}, \bibinfo {author} {\bibfnamefont {C.}~\bibnamefont {Zhou}}, \bibinfo
  {author} {\bibfnamefont {H.}~\bibnamefont {Zhou}}, \bibinfo {author}
  {\bibfnamefont {L.}~\bibnamefont {Han}}, \bibinfo {author} {\bibfnamefont
  {R.}~\bibnamefont {Chi}}, \bibinfo {author} {\bibfnamefont {K.}~\bibnamefont
  {Li}}, \bibinfo {author} {\bibfnamefont {M.}~\bibnamefont {Kofu}}, \bibinfo
  {author} {\bibfnamefont {K.}~\bibnamefont {Nakajima}}, \bibinfo {author}
  {\bibfnamefont {Y.}~\bibnamefont {Wei}}, \bibinfo {author} {\bibfnamefont
  {W.}~\bibnamefont {Zhang}}, \bibinfo {author} {\bibfnamefont {D.~G.}\
  \bibnamefont {Mazzone}}, \bibinfo {author} {\bibfnamefont {Z.~Y.}\
  \bibnamefont {Meng}},\ and\ \bibinfo {author} {\bibfnamefont
  {S.}~\bibnamefont {Li}},\ }\bibfield  {title} {\bibinfo {title} {{Spectral
  evidence for {Dirac} spinons in a kagome lattice antiferromagnet}},\ }\href
  {https://doi.org/10.1038/s41567-024-02495-z} {\bibfield  {journal} {\bibinfo
  {journal} {Nat. Phys.}\ }\textbf {\bibinfo {volume} {20}},\ \bibinfo {pages}
  {1097} (\bibinfo {year} {2024})}\BibitemShut {NoStop}%
\bibitem [{\citenamefont {Breidenbach}\ \emph {et~al.}(2025)\citenamefont
  {Breidenbach}, \citenamefont {Campello}, \citenamefont {Wen}, \citenamefont
  {Jiang}, \citenamefont {Pajerowski}, \citenamefont {Smaha},\ and\
  \citenamefont {Lee}}]{breidenbach2025}%
  \BibitemOpen
  \bibfield  {author} {\bibinfo {author} {\bibfnamefont {A.~T.}\ \bibnamefont
  {Breidenbach}}, \bibinfo {author} {\bibfnamefont {A.~C.}\ \bibnamefont
  {Campello}}, \bibinfo {author} {\bibfnamefont {J.}~\bibnamefont {Wen}},
  \bibinfo {author} {\bibfnamefont {H.-C.}\ \bibnamefont {Jiang}}, \bibinfo
  {author} {\bibfnamefont {D.~M.}\ \bibnamefont {Pajerowski}}, \bibinfo
  {author} {\bibfnamefont {R.~W.}\ \bibnamefont {Smaha}},\ and\ \bibinfo
  {author} {\bibfnamefont {Y.~S.}\ \bibnamefont {Lee}},\ }\bibfield  {title}
  {\bibinfo {title} {{Identifying universal spin excitations in candidate
  spin-1/2 kagome quantum spin liquid materials}},\ }\href
  {https://doi.org/10.1038/s41567-025-03069-3} {\bibfield  {journal} {\bibinfo
  {journal} {Nat. Phys.}\ }\textbf {\bibinfo {volume} {21}},\ \bibinfo {pages}
  {1957} (\bibinfo {year} {2025})}\BibitemShut {NoStop}%
\bibitem [{\citenamefont {Cui}\ \emph {et~al.}(2023)\citenamefont {Cui},
  \citenamefont {Liu}, \citenamefont {Lin}, \citenamefont {Wu}, \citenamefont
  {Hong}, \citenamefont {Liu}, \citenamefont {Li}, \citenamefont {Hu},
  \citenamefont {Xi}, \citenamefont {Li}, \citenamefont {Yu}, \citenamefont
  {Sandvik},\ and\ \citenamefont {Yu}}]{science.adc9487}%
  \BibitemOpen
  \bibfield  {author} {\bibinfo {author} {\bibfnamefont {Y.}~\bibnamefont
  {Cui}}, \bibinfo {author} {\bibfnamefont {L.}~\bibnamefont {Liu}}, \bibinfo
  {author} {\bibfnamefont {H.}~\bibnamefont {Lin}}, \bibinfo {author}
  {\bibfnamefont {K.-H.}\ \bibnamefont {Wu}}, \bibinfo {author} {\bibfnamefont
  {W.}~\bibnamefont {Hong}}, \bibinfo {author} {\bibfnamefont {X.}~\bibnamefont
  {Liu}}, \bibinfo {author} {\bibfnamefont {C.}~\bibnamefont {Li}}, \bibinfo
  {author} {\bibfnamefont {Z.}~\bibnamefont {Hu}}, \bibinfo {author}
  {\bibfnamefont {N.}~\bibnamefont {Xi}}, \bibinfo {author} {\bibfnamefont
  {S.}~\bibnamefont {Li}}, \bibinfo {author} {\bibfnamefont {R.}~\bibnamefont
  {Yu}}, \bibinfo {author} {\bibfnamefont {A.~W.}\ \bibnamefont {Sandvik}},\
  and\ \bibinfo {author} {\bibfnamefont {W.}~\bibnamefont {Yu}},\ }\bibfield
  {title} {\bibinfo {title} {{Proximate deconfined quantum critical point in
  SrCu$_2$(BO$_3$)$_2$}},\ }\href {https://doi.org/10.1126/science.adc9487}
  {\bibfield  {journal} {\bibinfo  {journal} {Science}\ }\textbf {\bibinfo
  {volume} {380}},\ \bibinfo {pages} {1179} (\bibinfo {year}
  {2023})}\BibitemShut {NoStop}%
\bibitem [{\citenamefont {Sandvik}(2026)}]{Sandvik_2026}%
  \BibitemOpen
  \bibfield  {author} {\bibinfo {author} {\bibfnamefont {A.~W.}\ \bibnamefont
  {Sandvik}},\ }\bibfield  {title} {\bibinfo {title} {High-precision ground
  state parameters of the two-dimensional spin-1/2 {Heisenberg} model on the
  square lattice},\ }\href {https://doi.org/10.1088/1742-5468/ae4f80}
  {\bibfield  {journal} {\bibinfo  {journal} {Journal of Statistical Mechanics:
  Theory and Experiment}\ }\textbf {\bibinfo {volume} {2026}},\ \bibinfo
  {pages} {043101} (\bibinfo {year} {2026})}\BibitemShut {NoStop}%
\bibitem [{\citenamefont {Karthik}\ and\ \citenamefont
  {Narayanan}(2024)}]{PhysRevD.109.034507}%
  \BibitemOpen
  \bibfield  {author} {\bibinfo {author} {\bibfnamefont {N.}~\bibnamefont
  {Karthik}}\ and\ \bibinfo {author} {\bibfnamefont {R.}~\bibnamefont
  {Narayanan}},\ }\bibfield  {title} {\bibinfo {title} {{Scaling dimension of
  the $4\ensuremath{\pi}$-flux monopole operator in four-flavor
  three-dimensional QED using lattice simulation}},\ }\href
  {https://doi.org/10.1103/PhysRevD.109.034507} {\bibfield  {journal} {\bibinfo
   {journal} {Phys. Rev. D}\ }\textbf {\bibinfo {volume} {109}},\ \bibinfo
  {pages} {034507} (\bibinfo {year} {2024})}\BibitemShut {NoStop}%
\bibitem [{\citenamefont {Verstraete}\ and\ \citenamefont
  {Cirac}(2004)}]{verstraete2004}%
  \BibitemOpen
  \bibfield  {author} {\bibinfo {author} {\bibfnamefont {F.}~\bibnamefont
  {Verstraete}}\ and\ \bibinfo {author} {\bibfnamefont {J.~I.}\ \bibnamefont
  {Cirac}},\ }\href {https://arxiv.org/abs/cond-mat/0407066} {\bibinfo {title}
  {{Renormalization algorithms for Quantum-Many Body Systems in two and higher
  dimensions}}} (\bibinfo {year} {2004}),\ \Eprint
  {https://arxiv.org/abs/cond-mat/0407066} {arXiv:cond-mat/0407066
  [cond-mat.str-el]} \BibitemShut {NoStop}%
\bibitem [{\citenamefont {Nishio}\ \emph {et~al.}(2004)\citenamefont {Nishio},
  \citenamefont {Maeshima}, \citenamefont {Gendiar},\ and\ \citenamefont
  {Nishino}}]{nishio2004tensor}%
  \BibitemOpen
  \bibfield  {author} {\bibinfo {author} {\bibfnamefont {Y.}~\bibnamefont
  {Nishio}}, \bibinfo {author} {\bibfnamefont {N.}~\bibnamefont {Maeshima}},
  \bibinfo {author} {\bibfnamefont {A.}~\bibnamefont {Gendiar}},\ and\ \bibinfo
  {author} {\bibfnamefont {T.}~\bibnamefont {Nishino}},\ }\bibfield  {title}
  {\bibinfo {title} {{Tensor product variational formulation for quantum
  systems}},\ }\href@noop {} {\bibfield  {journal} {\bibinfo  {journal} {arXiv
  preprint cond-mat/0401115}\ } (\bibinfo {year} {2004})}\BibitemShut {NoStop}%
\bibitem [{\citenamefont {Chen}\ and\ \citenamefont
  {Heyl}(2024)}]{Chen2024MinSR}%
  \BibitemOpen
  \bibfield  {author} {\bibinfo {author} {\bibfnamefont {A.}~\bibnamefont
  {Chen}}\ and\ \bibinfo {author} {\bibfnamefont {M.}~\bibnamefont {Heyl}},\
  }\bibfield  {title} {\bibinfo {title} {{Empowering deep neural quantum states
  through efficient optimization}},\ }\href
  {https://doi.org/10.1038/s41567-024-02566-1} {\bibfield  {journal} {\bibinfo
  {journal} {Nat. Phys.}\ }\textbf {\bibinfo {volume} {20}},\ \bibinfo {pages}
  {1476} (\bibinfo {year} {2024})}\BibitemShut {NoStop}%
\bibitem [{\citenamefont {Nomura}\ and\ \citenamefont
  {Imada}(2021)}]{nomuraDiracTypeNodalSpin2021a}%
  \BibitemOpen
  \bibfield  {author} {\bibinfo {author} {\bibfnamefont {Y.}~\bibnamefont
  {Nomura}}\ and\ \bibinfo {author} {\bibfnamefont {M.}~\bibnamefont {Imada}},\
  }\bibfield  {title} {\bibinfo {title} {{Dirac-Type Nodal Spin Liquid Revealed
  by Refined Quantum Many-Body Solver Using Neural-Network Wave Function,
  Correlation Ratio, and Level Spectroscopy}},\ }\href
  {https://doi.org/10.1103/PhysRevX.11.031034} {\bibfield  {journal} {\bibinfo
  {journal} {Phys. Rev. X}\ }\textbf {\bibinfo {volume} {11}},\ \bibinfo
  {pages} {031034} (\bibinfo {year} {2021})}\BibitemShut {NoStop}%
\bibitem [{\citenamefont {Viteritti}\ \emph {et~al.}(2026)\citenamefont
  {Viteritti}, \citenamefont {Rende}, \citenamefont {Sachdev},\ and\
  \citenamefont {Carleo}}]{Viteritti2026ThermodynamicNQS}%
  \BibitemOpen
  \bibfield  {author} {\bibinfo {author} {\bibfnamefont {L.~L.}\ \bibnamefont
  {Viteritti}}, \bibinfo {author} {\bibfnamefont {R.}~\bibnamefont {Rende}},
  \bibinfo {author} {\bibfnamefont {S.}~\bibnamefont {Sachdev}},\ and\ \bibinfo
  {author} {\bibfnamefont {G.}~\bibnamefont {Carleo}},\ }\href
  {https://arxiv.org/abs/2602.02665} {\bibinfo {title} {{Approaching the
  Thermodynamic Limit with Neural-Network Quantum States}}} (\bibinfo {year}
  {2026}),\ \Eprint {https://arxiv.org/abs/2602.02665} {arXiv:2602.02665
  [cond-mat.str-el]} \BibitemShut {NoStop}%
\bibitem [{\citenamefont {Fan}\ \emph {et~al.}(2026)\citenamefont {Fan},
  \citenamefont {Zhan}, \citenamefont {Gu}, \citenamefont {Liu}, \citenamefont
  {Wu}, \citenamefont {Qin}, \citenamefont {Lv},\ and\ \citenamefont
  {Xiang}}]{Fan2026NuTNS}%
  \BibitemOpen
  \bibfield  {author} {\bibinfo {author} {\bibfnamefont {C.}~\bibnamefont
  {Fan}}, \bibinfo {author} {\bibfnamefont {B.}~\bibnamefont {Zhan}}, \bibinfo
  {author} {\bibfnamefont {Y.}~\bibnamefont {Gu}}, \bibinfo {author}
  {\bibfnamefont {T.}~\bibnamefont {Liu}}, \bibinfo {author} {\bibfnamefont
  {Y.}~\bibnamefont {Wu}}, \bibinfo {author} {\bibfnamefont {M.}~\bibnamefont
  {Qin}}, \bibinfo {author} {\bibfnamefont {D.}~\bibnamefont {Lv}},\ and\
  \bibinfo {author} {\bibfnamefont {T.}~\bibnamefont {Xiang}},\ }\href
  {https://arxiv.org/abs/2603.14425} {\bibinfo {title} {{Disentangling Tensor
  Network States with Deep Neural Network}}} (\bibinfo {year} {2026}),\ \Eprint
  {https://arxiv.org/abs/2603.14425} {arXiv:2603.14425 [cond-mat.str-el]}
  \BibitemShut {NoStop}%
\bibitem [{\citenamefont {Wen}(2002)}]{PhysRevB.65.165113}%
  \BibitemOpen
  \bibfield  {author} {\bibinfo {author} {\bibfnamefont {X.-G.}\ \bibnamefont
  {Wen}},\ }\bibfield  {title} {\bibinfo {title} {{Quantum orders and symmetric
  spin liquids}},\ }\href {https://doi.org/10.1103/PhysRevB.65.165113}
  {\bibfield  {journal} {\bibinfo  {journal} {Phys. Rev. B}\ }\textbf {\bibinfo
  {volume} {65}},\ \bibinfo {pages} {165113} (\bibinfo {year}
  {2002})}\BibitemShut {NoStop}%
\bibitem [{\citenamefont {Hu}\ \emph {et~al.}(2013)\citenamefont {Hu},
  \citenamefont {Becca}, \citenamefont {Parola},\ and\ \citenamefont
  {Sorella}}]{PhysRevB.88.060402}%
  \BibitemOpen
  \bibfield  {author} {\bibinfo {author} {\bibfnamefont {W.-J.}\ \bibnamefont
  {Hu}}, \bibinfo {author} {\bibfnamefont {F.}~\bibnamefont {Becca}}, \bibinfo
  {author} {\bibfnamefont {A.}~\bibnamefont {Parola}},\ and\ \bibinfo {author}
  {\bibfnamefont {S.}~\bibnamefont {Sorella}},\ }\bibfield  {title} {\bibinfo
  {title} {{Direct evidence for a gapless ${Z}_{2}$ spin liquid by frustrating
  {N\'eel} antiferromagnetism}},\ }\href
  {https://doi.org/10.1103/PhysRevB.88.060402} {\bibfield  {journal} {\bibinfo
  {journal} {Phys. Rev. B}\ }\textbf {\bibinfo {volume} {88}},\ \bibinfo
  {pages} {060402} (\bibinfo {year} {2013})}\BibitemShut {NoStop}%
\bibitem [{\citenamefont {Shackleton}\ \emph {et~al.}(2021)\citenamefont
  {Shackleton}, \citenamefont {Thomson},\ and\ \citenamefont
  {Sachdev}}]{PhysRevB.104.045110}%
  \BibitemOpen
  \bibfield  {author} {\bibinfo {author} {\bibfnamefont {L.}~\bibnamefont
  {Shackleton}}, \bibinfo {author} {\bibfnamefont {A.}~\bibnamefont
  {Thomson}},\ and\ \bibinfo {author} {\bibfnamefont {S.}~\bibnamefont
  {Sachdev}},\ }\bibfield  {title} {\bibinfo {title} {Deconfined criticality
  and a gapless {$\mathbb{Z}_2$} spin liquid in the square-lattice
  antiferromagnet},\ }\href {https://doi.org/10.1103/PhysRevB.104.045110}
  {\bibfield  {journal} {\bibinfo  {journal} {Phys. Rev. B}\ }\textbf {\bibinfo
  {volume} {104}},\ \bibinfo {pages} {045110} (\bibinfo {year}
  {2021})}\BibitemShut {NoStop}%
\bibitem [{\citenamefont {Wang}\ and\ \citenamefont
  {Sandvik}(2018)}]{wangCriticalLevelCrossings2018}%
  \BibitemOpen
  \bibfield  {author} {\bibinfo {author} {\bibfnamefont {L.}~\bibnamefont
  {Wang}}\ and\ \bibinfo {author} {\bibfnamefont {A.~W.}\ \bibnamefont
  {Sandvik}},\ }\bibfield  {title} {\bibinfo {title} {{Critical Level Crossings
  and Gapless Spin Liquid in the Square-Lattice Spin-$1/2$
  ${J}_{1}\text{\ensuremath{-}}{J}_{2}$ {Heisenberg} Antiferromagnet}},\ }\href
  {https://doi.org/10.1103/PhysRevLett.121.107202} {\bibfield  {journal}
  {\bibinfo  {journal} {Phys. Rev. Lett.}\ }\textbf {\bibinfo {volume} {121}},\
  \bibinfo {pages} {107202} (\bibinfo {year} {2018})}\BibitemShut {NoStop}%
\bibitem [{\citenamefont {Zhang}\ \emph {et~al.}(2024)\citenamefont {Zhang},
  \citenamefont {Chi}, \citenamefont {Liu},\ and\ \citenamefont
  {Wang}}]{PhysRevB.109.075129}%
  \BibitemOpen
  \bibfield  {author} {\bibinfo {author} {\bibfnamefont {X.-Y.}\ \bibnamefont
  {Zhang}}, \bibinfo {author} {\bibfnamefont {R.}~\bibnamefont {Chi}}, \bibinfo
  {author} {\bibfnamefont {Y.}~\bibnamefont {Liu}},\ and\ \bibinfo {author}
  {\bibfnamefont {L.}~\bibnamefont {Wang}},\ }\bibfield  {title} {\bibinfo
  {title} {{Two-dimensional excitation information by a matrix product state
  method on infinite helices}},\ }\href
  {https://doi.org/10.1103/PhysRevB.109.075129} {\bibfield  {journal} {\bibinfo
   {journal} {Phys. Rev. B}\ }\textbf {\bibinfo {volume} {109}},\ \bibinfo
  {pages} {075129} (\bibinfo {year} {2024})}\BibitemShut {NoStop}%
\bibitem [{\citenamefont {Qian}\ and\ \citenamefont
  {Qin}(2024)}]{qianAbsenceSpinLiquid2024}%
  \BibitemOpen
  \bibfield  {author} {\bibinfo {author} {\bibfnamefont {X.}~\bibnamefont
  {Qian}}\ and\ \bibinfo {author} {\bibfnamefont {M.}~\bibnamefont {Qin}},\
  }\bibfield  {title} {\bibinfo {title} {{Absence of spin liquid phase in the
  ${J}_{1}\text{\ensuremath{-}}{J}_{2}$ {Heisenberg} model on the square
  lattice}},\ }\href {https://doi.org/10.1103/PhysRevB.109.L161103} {\bibfield
  {journal} {\bibinfo  {journal} {Phys. Rev. B}\ }\textbf {\bibinfo {volume}
  {109}},\ \bibinfo {pages} {L161103} (\bibinfo {year} {2024})}\BibitemShut
  {NoStop}%
\bibitem [{\citenamefont {Gong}\ \emph {et~al.}(2014)\citenamefont {Gong},
  \citenamefont {Zhu}, \citenamefont {Sheng}, \citenamefont {Motrunich},\ and\
  \citenamefont {Fisher}}]{gongPlaquetteOrderedPhase2014}%
  \BibitemOpen
  \bibfield  {author} {\bibinfo {author} {\bibfnamefont {S.-S.}\ \bibnamefont
  {Gong}}, \bibinfo {author} {\bibfnamefont {W.}~\bibnamefont {Zhu}}, \bibinfo
  {author} {\bibfnamefont {D.~N.}\ \bibnamefont {Sheng}}, \bibinfo {author}
  {\bibfnamefont {O.~I.}\ \bibnamefont {Motrunich}},\ and\ \bibinfo {author}
  {\bibfnamefont {M.~P.~A.}\ \bibnamefont {Fisher}},\ }\bibfield  {title}
  {\bibinfo {title} {{Plaquette Ordered Phase and Quantum Phase Diagram in the
  Spin-$1/2$ ${J}_{1}\text{\ensuremath{-}}{J}_{2}$ Square {Heisenberg}
  Model}},\ }\href {https://doi.org/10.1103/PhysRevLett.113.027201} {\bibfield
  {journal} {\bibinfo  {journal} {Phys. Rev. Lett.}\ }\textbf {\bibinfo
  {volume} {113}},\ \bibinfo {pages} {027201} (\bibinfo {year}
  {2014})}\BibitemShut {NoStop}%
\bibitem [{\citenamefont {Liu}\ \emph {et~al.}(2022)\citenamefont {Liu},
  \citenamefont {Gong}, \citenamefont {Li}, \citenamefont {Poilblanc},
  \citenamefont {Chen},\ and\ \citenamefont {Gu}}]{liuGaplessQuantumSpin2022a}%
  \BibitemOpen
  \bibfield  {author} {\bibinfo {author} {\bibfnamefont {W.-Y.}\ \bibnamefont
  {Liu}}, \bibinfo {author} {\bibfnamefont {S.-S.}\ \bibnamefont {Gong}},
  \bibinfo {author} {\bibfnamefont {Y.-B.}\ \bibnamefont {Li}}, \bibinfo
  {author} {\bibfnamefont {D.}~\bibnamefont {Poilblanc}}, \bibinfo {author}
  {\bibfnamefont {W.-Q.}\ \bibnamefont {Chen}},\ and\ \bibinfo {author}
  {\bibfnamefont {Z.-C.}\ \bibnamefont {Gu}},\ }\bibfield  {title} {\bibinfo
  {title} {{Gapless quantum spin liquid and global phase diagram of the
  spin-$1/2$ ${J}_{1}\text{\ensuremath{-}}{J}_{2}$ square antiferromagnetic
  {Heisenberg} model}},\ }\href {https://doi.org/10.1016/j.scib.2022.03.010}
  {\bibfield  {journal} {\bibinfo  {journal} {Sci. Bull.}\ }\textbf {\bibinfo
  {volume} {67}},\ \bibinfo {pages} {1034} (\bibinfo {year}
  {2022})}\BibitemShut {NoStop}%
\bibitem [{\citenamefont {Mambrini}\ \emph {et~al.}(2006)\citenamefont
  {Mambrini}, \citenamefont {L\"auchli}, \citenamefont {Poilblanc},\ and\
  \citenamefont {Mila}}]{PhysRevB.74.144422}%
  \BibitemOpen
  \bibfield  {author} {\bibinfo {author} {\bibfnamefont {M.}~\bibnamefont
  {Mambrini}}, \bibinfo {author} {\bibfnamefont {A.}~\bibnamefont {L\"auchli}},
  \bibinfo {author} {\bibfnamefont {D.}~\bibnamefont {Poilblanc}},\ and\
  \bibinfo {author} {\bibfnamefont {F.}~\bibnamefont {Mila}},\ }\bibfield
  {title} {\bibinfo {title} {{Plaquette valence-bond crystal in the frustrated
  Heisenberg quantum antiferromagnet on the square lattice}},\ }\href
  {https://doi.org/10.1103/PhysRevB.74.144422} {\bibfield  {journal} {\bibinfo
  {journal} {Phys. Rev. B}\ }\textbf {\bibinfo {volume} {74}},\ \bibinfo
  {pages} {144422} (\bibinfo {year} {2006})}\BibitemShut {NoStop}%
\bibitem [{\citenamefont {Huang}\ \emph {et~al.}(2024)\citenamefont {Huang},
  \citenamefont {Qian},\ and\ \citenamefont {Qin}}]{PhysRevB.110.195111}%
  \BibitemOpen
  \bibfield  {author} {\bibinfo {author} {\bibfnamefont {J.}~\bibnamefont
  {Huang}}, \bibinfo {author} {\bibfnamefont {X.}~\bibnamefont {Qian}},\ and\
  \bibinfo {author} {\bibfnamefont {M.}~\bibnamefont {Qin}},\ }\bibfield
  {title} {\bibinfo {title} {{Plaquette-type valence bond solid state in the
  ${J}_{1}\text{\ensuremath{-}}{J}_{2}$ square-lattice Heisenberg model}},\
  }\href {https://doi.org/10.1103/PhysRevB.110.195111} {\bibfield  {journal}
  {\bibinfo  {journal} {Phys. Rev. B}\ }\textbf {\bibinfo {volume} {110}},\
  \bibinfo {pages} {195111} (\bibinfo {year} {2024})}\BibitemShut {NoStop}%
\bibitem [{\citenamefont {Zhang}\ \emph
  {et~al.}(2026{\natexlab{a}})\citenamefont {Zhang}, \citenamefont {Yang},\
  and\ \citenamefont {Corboz}}]{Zhang2026QRCTMRG}%
  \BibitemOpen
  \bibfield  {author} {\bibinfo {author} {\bibfnamefont {Y.}~\bibnamefont
  {Zhang}}, \bibinfo {author} {\bibfnamefont {Q.}~\bibnamefont {Yang}},\ and\
  \bibinfo {author} {\bibfnamefont {P.}~\bibnamefont {Corboz}},\ }\bibfield
  {title} {\bibinfo {title} {{Accelerating two-dimensional tensor network
  contractions using QR decompositions}},\ }\href
  {https://doi.org/10.1103/ptls-kr9z} {\bibfield  {journal} {\bibinfo
  {journal} {Phys. Rev. B}\ }\textbf {\bibinfo {volume} {113}},\ \bibinfo
  {pages} {L201106} (\bibinfo {year} {2026}{\natexlab{a}})}\BibitemShut
  {NoStop}%
\bibitem [{\citenamefont {Liu}\ \emph {et~al.}(2021)\citenamefont {Liu},
  \citenamefont {Huang}, \citenamefont {Gong},\ and\ \citenamefont
  {Gu}}]{Liu2021FinitePEPS}%
  \BibitemOpen
  \bibfield  {author} {\bibinfo {author} {\bibfnamefont {W.-Y.}\ \bibnamefont
  {Liu}}, \bibinfo {author} {\bibfnamefont {Y.-Z.}\ \bibnamefont {Huang}},
  \bibinfo {author} {\bibfnamefont {S.-S.}\ \bibnamefont {Gong}},\ and\
  \bibinfo {author} {\bibfnamefont {Z.-C.}\ \bibnamefont {Gu}},\ }\bibfield
  {title} {\bibinfo {title} {Accurate simulation for finite projected entangled
  pair states in two dimensions},\ }\href
  {https://doi.org/10.1103/PhysRevB.103.235155} {\bibfield  {journal} {\bibinfo
   {journal} {Phys. Rev. B}\ }\textbf {\bibinfo {volume} {103}},\ \bibinfo
  {pages} {235155} (\bibinfo {year} {2021})}\BibitemShut {NoStop}%
\bibitem [{\citenamefont {Dong}\ \emph {et~al.}(2025)\citenamefont {Dong},
  \citenamefont {Wang}, \citenamefont {Zhang}, \citenamefont {Zhang},\ and\
  \citenamefont {He}}]{Dong2024BeyondBoundaries}%
  \BibitemOpen
  \bibfield  {author} {\bibinfo {author} {\bibfnamefont {S.}~\bibnamefont
  {Dong}}, \bibinfo {author} {\bibfnamefont {C.}~\bibnamefont {Wang}}, \bibinfo
  {author} {\bibfnamefont {H.}~\bibnamefont {Zhang}}, \bibinfo {author}
  {\bibfnamefont {M.}~\bibnamefont {Zhang}},\ and\ \bibinfo {author}
  {\bibfnamefont {L.}~\bibnamefont {He}},\ }\bibfield  {title} {\bibinfo
  {title} {Efficient projected entangled pair states methods for periodic
  quantum systems},\ }\href {https://doi.org/10.1103/fvq2-msr4} {\bibfield
  {journal} {\bibinfo  {journal} {Phys. Rev. Lett.}\ }\textbf {\bibinfo
  {volume} {135}},\ \bibinfo {pages} {026501} (\bibinfo {year}
  {2025})}\BibitemShut {NoStop}%
\bibitem [{\citenamefont {Hastings}(2000)}]{PhysRevB.63.014413}%
  \BibitemOpen
  \bibfield  {author} {\bibinfo {author} {\bibfnamefont {M.~B.}\ \bibnamefont
  {Hastings}},\ }\bibfield  {title} {\bibinfo {title} {{Dirac structure, RVB,
  and Goldstone modes in the kagom\'e antiferromagnet}},\ }\href
  {https://doi.org/10.1103/PhysRevB.63.014413} {\bibfield  {journal} {\bibinfo
  {journal} {Phys. Rev. B}\ }\textbf {\bibinfo {volume} {63}},\ \bibinfo
  {pages} {014413} (\bibinfo {year} {2000})}\BibitemShut {NoStop}%
\bibitem [{\citenamefont {Iqbal}\ \emph {et~al.}(2013)\citenamefont {Iqbal},
  \citenamefont {Becca}, \citenamefont {Sorella},\ and\ \citenamefont
  {Poilblanc}}]{PhysRevB.87.060405}%
  \BibitemOpen
  \bibfield  {author} {\bibinfo {author} {\bibfnamefont {Y.}~\bibnamefont
  {Iqbal}}, \bibinfo {author} {\bibfnamefont {F.}~\bibnamefont {Becca}},
  \bibinfo {author} {\bibfnamefont {S.}~\bibnamefont {Sorella}},\ and\ \bibinfo
  {author} {\bibfnamefont {D.}~\bibnamefont {Poilblanc}},\ }\bibfield  {title}
  {\bibinfo {title} {{Gapless spin-liquid phase in the kagome
  spin-$\frac{1}{2}$ Heisenberg antiferromagnet}},\ }\href
  {https://doi.org/10.1103/PhysRevB.87.060405} {\bibfield  {journal} {\bibinfo
  {journal} {Phys. Rev. B}\ }\textbf {\bibinfo {volume} {87}},\ \bibinfo
  {pages} {060405(R)} (\bibinfo {year} {2013})}\BibitemShut {NoStop}%
\bibitem [{\citenamefont {Iqbal}\ \emph {et~al.}(2015)\citenamefont {Iqbal},
  \citenamefont {Poilblanc},\ and\ \citenamefont {Becca}}]{Iqbal2015}%
  \BibitemOpen
  \bibfield  {author} {\bibinfo {author} {\bibfnamefont {Y.}~\bibnamefont
  {Iqbal}}, \bibinfo {author} {\bibfnamefont {D.}~\bibnamefont {Poilblanc}},\
  and\ \bibinfo {author} {\bibfnamefont {F.}~\bibnamefont {Becca}},\ }\bibfield
   {title} {\bibinfo {title} {{Spin-$1/2$ {Heisenberg} ${J}_{1}$-${J}_{2}$
  Antiferromagnet on the {Kagome} Lattice}},\ }\href
  {https://doi.org/10.1103/PhysRevB.91.020402} {\bibfield  {journal} {\bibinfo
  {journal} {Phys. Rev. B}\ }\textbf {\bibinfo {volume} {91}},\ \bibinfo
  {pages} {020402} (\bibinfo {year} {2015})}\BibitemShut {NoStop}%
\bibitem [{\citenamefont {Zhu}\ \emph {et~al.}(2019)\citenamefont {Zhu},
  \citenamefont {Gong},\ and\ \citenamefont {Sheng}}]{pnas.1807840116}%
  \BibitemOpen
  \bibfield  {author} {\bibinfo {author} {\bibfnamefont {W.}~\bibnamefont
  {Zhu}}, \bibinfo {author} {\bibfnamefont {S.-S.}\ \bibnamefont {Gong}},\ and\
  \bibinfo {author} {\bibfnamefont {D.~N.}\ \bibnamefont {Sheng}},\ }\bibfield
  {title} {\bibinfo {title} {Identifying spinon excitations from dynamic
  structure factor of spin-1/2 {Heisenberg} antiferromagnet on the kagome
  lattice},\ }\href {https://doi.org/10.1073/pnas.1807840116} {\bibfield
  {journal} {\bibinfo  {journal} {Proceedings of the National Academy of
  Sciences}\ }\textbf {\bibinfo {volume} {116}},\ \bibinfo {pages} {5437}
  (\bibinfo {year} {2019})}\BibitemShut {NoStop}%
\bibitem [{\citenamefont {Liao}\ \emph {et~al.}(2017)\citenamefont {Liao},
  \citenamefont {Xie}, \citenamefont {Chen}, \citenamefont {Liu}, \citenamefont
  {Xie}, \citenamefont {Huang}, \citenamefont {Normand},\ and\ \citenamefont
  {Xiang}}]{PhysRevLett.118.137202}%
  \BibitemOpen
  \bibfield  {author} {\bibinfo {author} {\bibfnamefont {H.~J.}\ \bibnamefont
  {Liao}}, \bibinfo {author} {\bibfnamefont {Z.~Y.}\ \bibnamefont {Xie}},
  \bibinfo {author} {\bibfnamefont {J.}~\bibnamefont {Chen}}, \bibinfo {author}
  {\bibfnamefont {Z.~Y.}\ \bibnamefont {Liu}}, \bibinfo {author} {\bibfnamefont
  {H.~D.}\ \bibnamefont {Xie}}, \bibinfo {author} {\bibfnamefont {R.~Z.}\
  \bibnamefont {Huang}}, \bibinfo {author} {\bibfnamefont {B.}~\bibnamefont
  {Normand}},\ and\ \bibinfo {author} {\bibfnamefont {T.}~\bibnamefont
  {Xiang}},\ }\bibfield  {title} {\bibinfo {title} {{Gapless Spin-Liquid Ground
  State in the $S=1/2$ Kagome Antiferromagnet}},\ }\href
  {https://doi.org/10.1103/PhysRevLett.118.137202} {\bibfield  {journal}
  {\bibinfo  {journal} {Phys. Rev. Lett.}\ }\textbf {\bibinfo {volume} {118}},\
  \bibinfo {pages} {137202} (\bibinfo {year} {2017})}\BibitemShut {NoStop}%
\bibitem [{\citenamefont {Song}\ \emph {et~al.}(2019)\citenamefont {Song},
  \citenamefont {Wang}, \citenamefont {Vishwanath},\ and\ \citenamefont
  {He}}]{Song2019}%
  \BibitemOpen
  \bibfield  {author} {\bibinfo {author} {\bibfnamefont {X.-Y.}\ \bibnamefont
  {Song}}, \bibinfo {author} {\bibfnamefont {C.}~\bibnamefont {Wang}}, \bibinfo
  {author} {\bibfnamefont {A.}~\bibnamefont {Vishwanath}},\ and\ \bibinfo
  {author} {\bibfnamefont {Y.-C.}\ \bibnamefont {He}},\ }\bibfield  {title}
  {\bibinfo {title} {Unifying description of competing orders in
  two-dimensional quantum magnets},\ }\href
  {https://doi.org/10.1038/s41467-019-11727-3} {\bibfield  {journal} {\bibinfo
  {journal} {Nature Communications}\ }\textbf {\bibinfo {volume} {10}},\
  \bibinfo {pages} {4254} (\bibinfo {year} {2019})}\BibitemShut {NoStop}%
\bibitem [{\citenamefont {Yan}\ \emph {et~al.}(2011)\citenamefont {Yan},
  \citenamefont {Huse},\ and\ \citenamefont {White}}]{Yan2011SpinLiquidKagome}%
  \BibitemOpen
  \bibfield  {author} {\bibinfo {author} {\bibfnamefont {S.}~\bibnamefont
  {Yan}}, \bibinfo {author} {\bibfnamefont {D.~A.}\ \bibnamefont {Huse}},\ and\
  \bibinfo {author} {\bibfnamefont {S.~R.}\ \bibnamefont {White}},\ }\bibfield
  {title} {\bibinfo {title} {Spin-liquid ground state of the {$S$ = 1/2} kagome
  {Heisenberg} antiferromagnet},\ }\href
  {https://doi.org/10.1126/science.1201080} {\bibfield  {journal} {\bibinfo
  {journal} {Science}\ }\textbf {\bibinfo {volume} {332}},\ \bibinfo {pages}
  {1173} (\bibinfo {year} {2011})}\BibitemShut {NoStop}%
\bibitem [{\citenamefont {Hermele}\ \emph {et~al.}(2008)\citenamefont
  {Hermele}, \citenamefont {Ran}, \citenamefont {Lee},\ and\ \citenamefont
  {Wen}}]{PhysRevB.77.224413}%
  \BibitemOpen
  \bibfield  {author} {\bibinfo {author} {\bibfnamefont {M.}~\bibnamefont
  {Hermele}}, \bibinfo {author} {\bibfnamefont {Y.}~\bibnamefont {Ran}},
  \bibinfo {author} {\bibfnamefont {P.~A.}\ \bibnamefont {Lee}},\ and\ \bibinfo
  {author} {\bibfnamefont {X.-G.}\ \bibnamefont {Wen}},\ }\bibfield  {title}
  {\bibinfo {title} {{Properties of an algebraic spin liquid on the kagome
  lattice}},\ }\href {https://doi.org/10.1103/PhysRevB.77.224413} {\bibfield
  {journal} {\bibinfo  {journal} {Phys. Rev. B}\ }\textbf {\bibinfo {volume}
  {77}},\ \bibinfo {pages} {224413} (\bibinfo {year} {2008})}\BibitemShut
  {NoStop}%
\bibitem [{\citenamefont {\DJ{}uri\ifmmode~\acute{c}\else \'{c}\fi{}}\ \emph
  {et~al.}(2025)\citenamefont {\DJ{}uri\ifmmode~\acute{c}\else \'{c}\fi{}},
  \citenamefont {Chung}, \citenamefont {Yang},\ and\ \citenamefont
  {Sengupta}}]{PhysRevX.15.011047}%
  \BibitemOpen
  \bibfield  {author} {\bibinfo {author} {\bibfnamefont {T.}~\bibnamefont
  {\DJ{}uri\ifmmode~\acute{c}\else \'{c}\fi{}}}, \bibinfo {author}
  {\bibfnamefont {J.~H.}\ \bibnamefont {Chung}}, \bibinfo {author}
  {\bibfnamefont {B.}~\bibnamefont {Yang}},\ and\ \bibinfo {author}
  {\bibfnamefont {P.}~\bibnamefont {Sengupta}},\ }\bibfield  {title} {\bibinfo
  {title} {{Spin-$1/2$ Kagome Heisenberg Antiferromagnet: Machine Learning
  Discovery of the Spinon Pair-Density-Wave Ground State}},\ }\href
  {https://doi.org/10.1103/PhysRevX.15.011047} {\bibfield  {journal} {\bibinfo
  {journal} {Phys. Rev. X}\ }\textbf {\bibinfo {volume} {15}},\ \bibinfo
  {pages} {011047} (\bibinfo {year} {2025})}\BibitemShut {NoStop}%
\bibitem [{\citenamefont {He}\ \emph {et~al.}(2017)\citenamefont {He},
  \citenamefont {Zaletel}, \citenamefont {Oshikawa},\ and\ \citenamefont
  {Pollmann}}]{he2017}%
  \BibitemOpen
  \bibfield  {author} {\bibinfo {author} {\bibfnamefont {Y.-C.}\ \bibnamefont
  {He}}, \bibinfo {author} {\bibfnamefont {M.~P.}\ \bibnamefont {Zaletel}},
  \bibinfo {author} {\bibfnamefont {M.}~\bibnamefont {Oshikawa}},\ and\
  \bibinfo {author} {\bibfnamefont {F.}~\bibnamefont {Pollmann}},\ }\bibfield
  {title} {\bibinfo {title} {{Signatures of {Dirac} Cones in a {DMRG} Study of
  the {Kagome} {Heisenberg} Model}},\ }\href
  {https://doi.org/10.1103/PhysRevX.7.031020} {\bibfield  {journal} {\bibinfo
  {journal} {Phys. Rev. X}\ }\textbf {\bibinfo {volume} {7}},\ \bibinfo {pages}
  {031020} (\bibinfo {year} {2017})}\BibitemShut {NoStop}%
\bibitem [{\citenamefont {L\"auchli}\ \emph {et~al.}(2019)\citenamefont
  {L\"auchli}, \citenamefont {Sudan},\ and\ \citenamefont
  {Moessner}}]{PhysRevB.100.155142}%
  \BibitemOpen
  \bibfield  {author} {\bibinfo {author} {\bibfnamefont {A.~M.}\ \bibnamefont
  {L\"auchli}}, \bibinfo {author} {\bibfnamefont {J.}~\bibnamefont {Sudan}},\
  and\ \bibinfo {author} {\bibfnamefont {R.}~\bibnamefont {Moessner}},\
  }\bibfield  {title} {\bibinfo {title} {{$S=\frac{1}{2}$ kagome Heisenberg
  antiferromagnet revisited}},\ }\href
  {https://doi.org/10.1103/PhysRevB.100.155142} {\bibfield  {journal} {\bibinfo
   {journal} {Phys. Rev. B}\ }\textbf {\bibinfo {volume} {100}},\ \bibinfo
  {pages} {155142} (\bibinfo {year} {2019})}\BibitemShut {NoStop}%
\bibitem [{\citenamefont {Vanhecke}\ \emph {et~al.}(2022)\citenamefont
  {Vanhecke}, \citenamefont {Hasik}, \citenamefont {Verstraete},\ and\
  \citenamefont {Vanderstraeten}}]{PhysRevLett.129.200601}%
  \BibitemOpen
  \bibfield  {author} {\bibinfo {author} {\bibfnamefont {B.}~\bibnamefont
  {Vanhecke}}, \bibinfo {author} {\bibfnamefont {J.}~\bibnamefont {Hasik}},
  \bibinfo {author} {\bibfnamefont {F.}~\bibnamefont {Verstraete}},\ and\
  \bibinfo {author} {\bibfnamefont {L.}~\bibnamefont {Vanderstraeten}},\
  }\bibfield  {title} {\bibinfo {title} {{Scaling Hypothesis for Projected
  Entangled-Pair States}},\ }\href
  {https://doi.org/10.1103/PhysRevLett.129.200601} {\bibfield  {journal}
  {\bibinfo  {journal} {Phys. Rev. Lett.}\ }\textbf {\bibinfo {volume} {129}},\
  \bibinfo {pages} {200601} (\bibinfo {year} {2022})}\BibitemShut {NoStop}%
\bibitem [{\citenamefont {Rader}\ and\ \citenamefont
  {L\"auchli}(2018)}]{PhysRevX.8.031030}%
  \BibitemOpen
  \bibfield  {author} {\bibinfo {author} {\bibfnamefont {M.}~\bibnamefont
  {Rader}}\ and\ \bibinfo {author} {\bibfnamefont {A.~M.}\ \bibnamefont
  {L\"auchli}},\ }\bibfield  {title} {\bibinfo {title} {{Finite Correlation
  Length Scaling in Lorentz-Invariant Gapless iPEPS Wave Functions}},\ }\href
  {https://doi.org/10.1103/PhysRevX.8.031030} {\bibfield  {journal} {\bibinfo
  {journal} {Phys. Rev. X}\ }\textbf {\bibinfo {volume} {8}},\ \bibinfo {pages}
  {031030} (\bibinfo {year} {2018})}\BibitemShut {NoStop}%
\bibitem [{\citenamefont {Corboz}\ \emph {et~al.}(2018)\citenamefont {Corboz},
  \citenamefont {Czarnik}, \citenamefont {Kapteijns},\ and\ \citenamefont
  {Tagliacozzo}}]{PhysRevX.8.031031}%
  \BibitemOpen
  \bibfield  {author} {\bibinfo {author} {\bibfnamefont {P.}~\bibnamefont
  {Corboz}}, \bibinfo {author} {\bibfnamefont {P.}~\bibnamefont {Czarnik}},
  \bibinfo {author} {\bibfnamefont {G.}~\bibnamefont {Kapteijns}},\ and\
  \bibinfo {author} {\bibfnamefont {L.}~\bibnamefont {Tagliacozzo}},\
  }\bibfield  {title} {\bibinfo {title} {{Finite Correlation Length Scaling
  with Infinite Projected Entangled-Pair States}},\ }\href
  {https://doi.org/10.1103/PhysRevX.8.031031} {\bibfield  {journal} {\bibinfo
  {journal} {Phys. Rev. X}\ }\textbf {\bibinfo {volume} {8}},\ \bibinfo {pages}
  {031031} (\bibinfo {year} {2018})}\BibitemShut {NoStop}%
\bibitem [{\citenamefont {Hu}\ \emph {et~al.}(2026)\citenamefont {Hu},
  \citenamefont {Chi}, \citenamefont {Guo}, \citenamefont {Normand},
  \citenamefont {Liao},\ and\ \citenamefont
  {Xiang}}]{hu2026dynamicalspectralfunctionkagome}%
  \BibitemOpen
  \bibfield  {author} {\bibinfo {author} {\bibfnamefont {J.}~\bibnamefont
  {Hu}}, \bibinfo {author} {\bibfnamefont {R.}~\bibnamefont {Chi}}, \bibinfo
  {author} {\bibfnamefont {Y.}~\bibnamefont {Guo}}, \bibinfo {author}
  {\bibfnamefont {B.}~\bibnamefont {Normand}}, \bibinfo {author} {\bibfnamefont
  {H.-J.}\ \bibnamefont {Liao}},\ and\ \bibinfo {author} {\bibfnamefont
  {T.}~\bibnamefont {Xiang}},\ }\href {https://arxiv.org/abs/2512.18831}
  {\bibinfo {title} {{Dynamical Spectral Function of the Kagome Quantum Spin
  Liquid}}} (\bibinfo {year} {2026}),\ \Eprint
  {https://arxiv.org/abs/2512.18831} {arXiv:2512.18831 [cond-mat.str-el]}
  \BibitemShut {NoStop}%
\bibitem [{\citenamefont {Corboz}\ \emph {et~al.}(2012)\citenamefont {Corboz},
  \citenamefont {Penc}, \citenamefont {Mila},\ and\ \citenamefont
  {L\"auchli}}]{PhysRevB.86.041106}%
  \BibitemOpen
  \bibfield  {author} {\bibinfo {author} {\bibfnamefont {P.}~\bibnamefont
  {Corboz}}, \bibinfo {author} {\bibfnamefont {K.}~\bibnamefont {Penc}},
  \bibinfo {author} {\bibfnamefont {F.}~\bibnamefont {Mila}},\ and\ \bibinfo
  {author} {\bibfnamefont {A.~M.}\ \bibnamefont {L\"auchli}},\ }\bibfield
  {title} {\bibinfo {title} {{Simplex solids in SU($N$) Heisenberg models on
  the kagome and checkerboard lattices}},\ }\href
  {https://doi.org/10.1103/PhysRevB.86.041106} {\bibfield  {journal} {\bibinfo
  {journal} {Phys. Rev. B}\ }\textbf {\bibinfo {volume} {86}},\ \bibinfo
  {pages} {041106(R)} (\bibinfo {year} {2012})}\BibitemShut {NoStop}%
\bibitem [{\citenamefont {He}\ \emph {et~al.}(2021)\citenamefont {He},
  \citenamefont {Pekker},\ and\ \citenamefont {Mong}}]{PhysRevB.104.195126}%
  \BibitemOpen
  \bibfield  {author} {\bibinfo {author} {\bibfnamefont {Y.}~\bibnamefont
  {He}}, \bibinfo {author} {\bibfnamefont {D.}~\bibnamefont {Pekker}},\ and\
  \bibinfo {author} {\bibfnamefont {R.~S.~K.}\ \bibnamefont {Mong}},\
  }\bibfield  {title} {\bibinfo {title} {{One-dimensional repulsive Hubbard
  model with mass imbalance: Orders and filling anomaly}},\ }\href
  {https://doi.org/10.1103/PhysRevB.104.195126} {\bibfield  {journal} {\bibinfo
   {journal} {Phys. Rev. B}\ }\textbf {\bibinfo {volume} {104}},\ \bibinfo
  {pages} {195126} (\bibinfo {year} {2021})}\BibitemShut {NoStop}%
\bibitem [{\citenamefont {Liao}\ \emph {et~al.}(2019)\citenamefont {Liao},
  \citenamefont {Liu}, \citenamefont {Wang},\ and\ \citenamefont
  {Xiang}}]{ADPEPS}%
  \BibitemOpen
  \bibfield  {author} {\bibinfo {author} {\bibfnamefont {H.-J.}\ \bibnamefont
  {Liao}}, \bibinfo {author} {\bibfnamefont {J.-G.}\ \bibnamefont {Liu}},
  \bibinfo {author} {\bibfnamefont {L.}~\bibnamefont {Wang}},\ and\ \bibinfo
  {author} {\bibfnamefont {T.}~\bibnamefont {Xiang}},\ }\bibfield  {title}
  {\bibinfo {title} {{Differentiable Programming Tensor Networks}},\ }\href
  {https://doi.org/10.1103/PhysRevX.9.031041} {\bibfield  {journal} {\bibinfo
  {journal} {Phys. Rev. X}\ }\textbf {\bibinfo {volume} {9}},\ \bibinfo {pages}
  {031041} (\bibinfo {year} {2019})}\BibitemShut {NoStop}%
\bibitem [{\citenamefont {Corboz}(2016)}]{VPEPS}%
  \BibitemOpen
  \bibfield  {author} {\bibinfo {author} {\bibfnamefont {P.}~\bibnamefont
  {Corboz}},\ }\bibfield  {title} {\bibinfo {title} {{Variational optimization
  with infinite projected entangled-pair states}},\ }\href
  {https://doi.org/10.1103/PhysRevB.94.035133} {\bibfield  {journal} {\bibinfo
  {journal} {Phys. Rev. B}\ }\textbf {\bibinfo {volume} {94}},\ \bibinfo
  {pages} {035133} (\bibinfo {year} {2016})}\BibitemShut {NoStop}%
\bibitem [{\citenamefont {Francuz}\ \emph {et~al.}(2025)\citenamefont
  {Francuz}, \citenamefont {Schuch},\ and\ \citenamefont
  {Vanhecke}}]{PhysRevResearch.7.013237}%
  \BibitemOpen
  \bibfield  {author} {\bibinfo {author} {\bibfnamefont {A.}~\bibnamefont
  {Francuz}}, \bibinfo {author} {\bibfnamefont {N.}~\bibnamefont {Schuch}},\
  and\ \bibinfo {author} {\bibfnamefont {B.}~\bibnamefont {Vanhecke}},\
  }\bibfield  {title} {\bibinfo {title} {Stable and efficient differentiation
  of tensor network algorithms},\ }\href
  {https://doi.org/10.1103/PhysRevResearch.7.013237} {\bibfield  {journal}
  {\bibinfo  {journal} {Phys. Rev. Res.}\ }\textbf {\bibinfo {volume} {7}},\
  \bibinfo {pages} {013237} (\bibinfo {year} {2025})}\BibitemShut {NoStop}%
\bibitem [{\citenamefont {Burgelman}\ \emph {et~al.}(2026)\citenamefont
  {Burgelman}, \citenamefont {Francuz}, \citenamefont {Brehmer}, \citenamefont
  {Devos}, \citenamefont {Haegeman}, \citenamefont {Verstraete},\ and\
  \citenamefont
  {Vanhecke}}]{burgelman2026implicitdifferentiationtensornetwork}%
  \BibitemOpen
  \bibfield  {author} {\bibinfo {author} {\bibfnamefont {L.}~\bibnamefont
  {Burgelman}}, \bibinfo {author} {\bibfnamefont {A.}~\bibnamefont {Francuz}},
  \bibinfo {author} {\bibfnamefont {P.}~\bibnamefont {Brehmer}}, \bibinfo
  {author} {\bibfnamefont {L.}~\bibnamefont {Devos}}, \bibinfo {author}
  {\bibfnamefont {J.}~\bibnamefont {Haegeman}}, \bibinfo {author}
  {\bibfnamefont {F.}~\bibnamefont {Verstraete}},\ and\ \bibinfo {author}
  {\bibfnamefont {B.}~\bibnamefont {Vanhecke}},\ }\href
  {https://arxiv.org/abs/2607.15030} {\bibinfo {title} {{Implicit
  differentiation of tensor network algorithms}}} (\bibinfo {year} {2026}),\
  \Eprint {https://arxiv.org/abs/2607.15030} {arXiv:2607.15030 [quant-ph]}
  \BibitemShut {NoStop}%
\bibitem [{\citenamefont {Zhang}\ \emph
  {et~al.}(2026{\natexlab{b}})\citenamefont {Zhang}, \citenamefont {Yang},
  \citenamefont {Corboz}, \citenamefont {Haegeman},\ and\ \citenamefont
  {Tang}}]{h396-yc28}%
  \BibitemOpen
  \bibfield  {author} {\bibinfo {author} {\bibfnamefont {X.-Y.}\ \bibnamefont
  {Zhang}}, \bibinfo {author} {\bibfnamefont {Q.}~\bibnamefont {Yang}},
  \bibinfo {author} {\bibfnamefont {P.}~\bibnamefont {Corboz}}, \bibinfo
  {author} {\bibfnamefont {J.}~\bibnamefont {Haegeman}},\ and\ \bibinfo
  {author} {\bibfnamefont {W.}~\bibnamefont {Tang}},\ }\bibfield  {title}
  {\bibinfo {title} {{Accelerating two-dimensional tensor network optimization
  by preconditioning}},\ }\href {https://doi.org/10.1103/h396-yc28} {\bibfield
  {journal} {\bibinfo  {journal} {Phys. Rev. B}\ }\textbf {\bibinfo {volume}
  {113}},\ \bibinfo {pages} {125111} (\bibinfo {year}
  {2026}{\natexlab{b}})}\BibitemShut {NoStop}%
\bibitem [{\citenamefont {Vanderstraeten}\ \emph {et~al.}(2022)\citenamefont
  {Vanderstraeten}, \citenamefont {Burgelman}, \citenamefont {Ponsioen},
  \citenamefont {Van~Damme}, \citenamefont {Vanhecke}, \citenamefont {Corboz},
  \citenamefont {Haegeman},\ and\ \citenamefont
  {Verstraete}}]{PhysRevB.105.195140}%
  \BibitemOpen
  \bibfield  {author} {\bibinfo {author} {\bibfnamefont {L.}~\bibnamefont
  {Vanderstraeten}}, \bibinfo {author} {\bibfnamefont {L.}~\bibnamefont
  {Burgelman}}, \bibinfo {author} {\bibfnamefont {B.}~\bibnamefont {Ponsioen}},
  \bibinfo {author} {\bibfnamefont {M.}~\bibnamefont {Van~Damme}}, \bibinfo
  {author} {\bibfnamefont {B.}~\bibnamefont {Vanhecke}}, \bibinfo {author}
  {\bibfnamefont {P.}~\bibnamefont {Corboz}}, \bibinfo {author} {\bibfnamefont
  {J.}~\bibnamefont {Haegeman}},\ and\ \bibinfo {author} {\bibfnamefont
  {F.}~\bibnamefont {Verstraete}},\ }\bibfield  {title} {\bibinfo {title}
  {{Variational methods for contracting projected entangled-pair states}},\
  }\href {https://doi.org/10.1103/PhysRevB.105.195140} {\bibfield  {journal}
  {\bibinfo  {journal} {Phys. Rev. B}\ }\textbf {\bibinfo {volume} {105}},\
  \bibinfo {pages} {195140} (\bibinfo {year} {2022})}\BibitemShut {NoStop}%
\bibitem [{\citenamefont {Acuaviva}\ \emph {et~al.}(2023)\citenamefont
  {Acuaviva}, \citenamefont {Makam}, \citenamefont {Nieuwboer}, \citenamefont
  {P\'erez-Garc\'ia}, \citenamefont {Sittner}, \citenamefont {Walter},\ and\
  \citenamefont {Witteveen}}]{Acuaviva2023MinimalCanonicalForm}%
  \BibitemOpen
  \bibfield  {author} {\bibinfo {author} {\bibfnamefont {A.}~\bibnamefont
  {Acuaviva}}, \bibinfo {author} {\bibfnamefont {V.}~\bibnamefont {Makam}},
  \bibinfo {author} {\bibfnamefont {H.}~\bibnamefont {Nieuwboer}}, \bibinfo
  {author} {\bibfnamefont {D.}~\bibnamefont {P\'erez-Garc\'ia}}, \bibinfo
  {author} {\bibfnamefont {F.}~\bibnamefont {Sittner}}, \bibinfo {author}
  {\bibfnamefont {M.}~\bibnamefont {Walter}},\ and\ \bibinfo {author}
  {\bibfnamefont {F.}~\bibnamefont {Witteveen}},\ }\bibfield  {title} {\bibinfo
  {title} {{The minimal canonical form of a tensor network}},\ }in\ \href
  {https://doi.org/10.1109/FOCS57990.2023.00027} {\emph {\bibinfo {booktitle}
  {2023 IEEE 64th Annual Symposium on Foundations of Computer Science
  (FOCS)}}}\ (\bibinfo  {publisher} {IEEE},\ \bibinfo {year} {2023})\ pp.\
  \bibinfo {pages} {328--362}\BibitemShut {NoStop}%
\bibitem [{\citenamefont {Tang}\ \emph {et~al.}(2025)\citenamefont {Tang},
  \citenamefont {Vanderstraeten},\ and\ \citenamefont
  {Haegeman}}]{Tang2025Gauging}%
  \BibitemOpen
  \bibfield  {author} {\bibinfo {author} {\bibfnamefont {W.}~\bibnamefont
  {Tang}}, \bibinfo {author} {\bibfnamefont {L.}~\bibnamefont
  {Vanderstraeten}},\ and\ \bibinfo {author} {\bibfnamefont {J.}~\bibnamefont
  {Haegeman}},\ }\href {https://arxiv.org/abs/2508.10822} {\bibinfo {title}
  {{Gauging the variational optimization of projected entangled-pair states}}}
  (\bibinfo {year} {2025}),\ \Eprint {https://arxiv.org/abs/2508.10822}
  {arXiv:2508.10822 [cond-mat.str-el]} \BibitemShut {NoStop}%
\bibitem [{\citenamefont {Zauner-Stauber}\ \emph {et~al.}(2018)\citenamefont
  {Zauner-Stauber}, \citenamefont {Vanderstraeten}, \citenamefont {Fishman},
  \citenamefont {Verstraete},\ and\ \citenamefont
  {Haegeman}}]{PhysRevB.97.045145}%
  \BibitemOpen
  \bibfield  {author} {\bibinfo {author} {\bibfnamefont {V.}~\bibnamefont
  {Zauner-Stauber}}, \bibinfo {author} {\bibfnamefont {L.}~\bibnamefont
  {Vanderstraeten}}, \bibinfo {author} {\bibfnamefont {M.~T.}\ \bibnamefont
  {Fishman}}, \bibinfo {author} {\bibfnamefont {F.}~\bibnamefont
  {Verstraete}},\ and\ \bibinfo {author} {\bibfnamefont {J.}~\bibnamefont
  {Haegeman}},\ }\bibfield  {title} {\bibinfo {title} {{Variational
  optimization algorithms for uniform matrix product states}},\ }\href
  {https://doi.org/10.1103/PhysRevB.97.045145} {\bibfield  {journal} {\bibinfo
  {journal} {Phys. Rev. B}\ }\textbf {\bibinfo {volume} {97}},\ \bibinfo
  {pages} {045145} (\bibinfo {year} {2018})}\BibitemShut {NoStop}%
\bibitem [{\citenamefont {Nietner}\ \emph {et~al.}(2020)\citenamefont
  {Nietner}, \citenamefont {Vanhecke}, \citenamefont {Verstraete},
  \citenamefont {Eisert},\ and\ \citenamefont
  {Vanderstraeten}}]{Nietner2020efficient}%
  \BibitemOpen
  \bibfield  {author} {\bibinfo {author} {\bibfnamefont {A.}~\bibnamefont
  {Nietner}}, \bibinfo {author} {\bibfnamefont {B.}~\bibnamefont {Vanhecke}},
  \bibinfo {author} {\bibfnamefont {F.}~\bibnamefont {Verstraete}}, \bibinfo
  {author} {\bibfnamefont {J.}~\bibnamefont {Eisert}},\ and\ \bibinfo {author}
  {\bibfnamefont {L.}~\bibnamefont {Vanderstraeten}},\ }\bibfield  {title}
  {\bibinfo {title} {{Efficient variational contraction of two-dimensional
  tensor networks with a non-trivial unit cell}},\ }\href
  {https://doi.org/10.22331/q-2020-09-21-328} {\bibfield  {journal} {\bibinfo
  {journal} {Quantum}\ }\textbf {\bibinfo {volume} {4}},\ \bibinfo {pages}
  {328} (\bibinfo {year} {2020})}\BibitemShut {NoStop}%
\bibitem [{\citenamefont {Zhang}\ \emph {et~al.}(2023)\citenamefont {Zhang},
  \citenamefont {Liang}, \citenamefont {Liao}, \citenamefont {Li},\ and\
  \citenamefont {Wang}}]{PhysRevB.108.085103}%
  \BibitemOpen
  \bibfield  {author} {\bibinfo {author} {\bibfnamefont {X.-Y.}\ \bibnamefont
  {Zhang}}, \bibinfo {author} {\bibfnamefont {S.}~\bibnamefont {Liang}},
  \bibinfo {author} {\bibfnamefont {H.-J.}\ \bibnamefont {Liao}}, \bibinfo
  {author} {\bibfnamefont {W.}~\bibnamefont {Li}},\ and\ \bibinfo {author}
  {\bibfnamefont {L.}~\bibnamefont {Wang}},\ }\bibfield  {title} {\bibinfo
  {title} {{Differentiable programming tensor networks for Kitaev magnets}},\
  }\href {https://doi.org/10.1103/PhysRevB.108.085103} {\bibfield  {journal}
  {\bibinfo  {journal} {Phys. Rev. B}\ }\textbf {\bibinfo {volume} {108}},\
  \bibinfo {pages} {085103} (\bibinfo {year} {2023})}\BibitemShut {NoStop}%
\bibitem [{\citenamefont {Zhang}\ \emph
  {et~al.}(2026{\natexlab{c}})\citenamefont {Zhang}, \citenamefont {Yang},
  \citenamefont {Corboz}, \citenamefont {Haegeman},\ and\ \citenamefont
  {He}}]{Zhang2025HigherSpinKitaev}%
  \BibitemOpen
  \bibfield  {author} {\bibinfo {author} {\bibfnamefont {X.-Y.}\ \bibnamefont
  {Zhang}}, \bibinfo {author} {\bibfnamefont {Q.}~\bibnamefont {Yang}},
  \bibinfo {author} {\bibfnamefont {P.}~\bibnamefont {Corboz}}, \bibinfo
  {author} {\bibfnamefont {J.}~\bibnamefont {Haegeman}},\ and\ \bibinfo
  {author} {\bibfnamefont {Y.}~\bibnamefont {He}},\ }\bibfield  {title}
  {\bibinfo {title} {Topological and trivial valence-bond orders in higher-spin
  {Kitaev} models},\ }\href {https://doi.org/10.1103/7q31-xxsd} {\bibfield
  {journal} {\bibinfo  {journal} {Phys. Rev. Lett.}\ }\textbf {\bibinfo
  {volume} {137}},\ \bibinfo {pages} {176502} (\bibinfo {year}
  {2026}{\natexlab{c}})}\BibitemShut {NoStop}%
\bibitem [{\citenamefont {Nishino}\ and\ \citenamefont {Okunishi}(1996)}]{CTM}%
  \BibitemOpen
  \bibfield  {author} {\bibinfo {author} {\bibfnamefont {T.}~\bibnamefont
  {Nishino}}\ and\ \bibinfo {author} {\bibfnamefont {K.}~\bibnamefont
  {Okunishi}},\ }\bibfield  {title} {\bibinfo {title} {{Corner Transfer Matrix
  Renormalization Group Method}},\ }\href {https://doi.org/10.1143/JPSJ.65.891}
  {\bibfield  {journal} {\bibinfo  {journal} {J. Phys. Soc. Jpn.}\ }\textbf
  {\bibinfo {volume} {65}},\ \bibinfo {pages} {891} (\bibinfo {year}
  {1996})}\BibitemShut {NoStop}%
\bibitem [{\citenamefont {Corboz}\ \emph {et~al.}(2014)\citenamefont {Corboz},
  \citenamefont {Rice},\ and\ \citenamefont {Troyer}}]{PhysRevLett.113.046402}%
  \BibitemOpen
  \bibfield  {author} {\bibinfo {author} {\bibfnamefont {P.}~\bibnamefont
  {Corboz}}, \bibinfo {author} {\bibfnamefont {T.~M.}\ \bibnamefont {Rice}},\
  and\ \bibinfo {author} {\bibfnamefont {M.}~\bibnamefont {Troyer}},\
  }\bibfield  {title} {\bibinfo {title} {{Competing States in the $t$-$J$
  Model: Uniform $d$-Wave State versus Stripe State}},\ }\href
  {https://doi.org/10.1103/PhysRevLett.113.046402} {\bibfield  {journal}
  {\bibinfo  {journal} {Phys. Rev. Lett.}\ }\textbf {\bibinfo {volume} {113}},\
  \bibinfo {pages} {046402} (\bibinfo {year} {2014})}\BibitemShut {NoStop}%
\bibitem [{\citenamefont {Lukin}\ and\ \citenamefont
  {Sotnikov}(2023)}]{PhysRevB.107.054424}%
  \BibitemOpen
  \bibfield  {author} {\bibinfo {author} {\bibfnamefont {I.~V.}\ \bibnamefont
  {Lukin}}\ and\ \bibinfo {author} {\bibfnamefont {A.~G.}\ \bibnamefont
  {Sotnikov}},\ }\bibfield  {title} {\bibinfo {title} {{Variational
  optimization of tensor-network states with the honeycomb-lattice corner
  transfer matrix}},\ }\href {https://doi.org/10.1103/PhysRevB.107.054424}
  {\bibfield  {journal} {\bibinfo  {journal} {Phys. Rev. B}\ }\textbf {\bibinfo
  {volume} {107}},\ \bibinfo {pages} {054424} (\bibinfo {year}
  {2023})}\BibitemShut {NoStop}%
\bibitem [{\citenamefont {Richards}\ and\ \citenamefont
  {Sørensen}(2025)}]{richards2025acetngpuacceleratedcornertransfermatrixrenormalization}%
  \BibitemOpen
  \bibfield  {author} {\bibinfo {author} {\bibfnamefont {A.~D.~S.}\
  \bibnamefont {Richards}}\ and\ \bibinfo {author} {\bibfnamefont {E.~S.}\
  \bibnamefont {Sørensen}},\ }\href {https://arxiv.org/abs/2503.13900}
  {\bibinfo {title} {{ACE-TN}: {GPU}-accelerated corner-transfer-matrix
  renormalization of infinite projected entangled-pair states}} (\bibinfo
  {year} {2025}),\ \Eprint {https://arxiv.org/abs/2503.13900} {arXiv:2503.13900
  [cond-mat.str-el]} \BibitemShut {NoStop}%
\bibitem [{\citenamefont {Naumann}\ \emph {et~al.}(2025)\citenamefont
  {Naumann}, \citenamefont {Weerda}, \citenamefont {Eisert}, \citenamefont
  {Rizzi},\ and\ \citenamefont {Schmoll}}]{PhysRevB.111.235116}%
  \BibitemOpen
  \bibfield  {author} {\bibinfo {author} {\bibfnamefont {J.}~\bibnamefont
  {Naumann}}, \bibinfo {author} {\bibfnamefont {E.~L.}\ \bibnamefont {Weerda}},
  \bibinfo {author} {\bibfnamefont {J.}~\bibnamefont {Eisert}}, \bibinfo
  {author} {\bibfnamefont {M.}~\bibnamefont {Rizzi}},\ and\ \bibinfo {author}
  {\bibfnamefont {P.}~\bibnamefont {Schmoll}},\ }\bibfield  {title} {\bibinfo
  {title} {{Variationally optimizing infinite projected entangled-pair states
  at large bond dimensions: A split corner transfer matrix renormalization
  group approach}},\ }\href {https://doi.org/10.1103/PhysRevB.111.235116}
  {\bibfield  {journal} {\bibinfo  {journal} {Phys. Rev. B}\ }\textbf {\bibinfo
  {volume} {111}},\ \bibinfo {pages} {235116} (\bibinfo {year}
  {2025})}\BibitemShut {NoStop}%
\bibitem [{\citenamefont {Zhang}\ and\ \citenamefont
  {Corboz}(2026)}]{zhang2026fasttwodimensionaltensornetworkcontraction}%
  \BibitemOpen
  \bibfield  {author} {\bibinfo {author} {\bibfnamefont {Y.}~\bibnamefont
  {Zhang}}\ and\ \bibinfo {author} {\bibfnamefont {P.}~\bibnamefont {Corboz}},\
  }\href {https://arxiv.org/abs/2607.15158} {\bibinfo {title} {Fast
  two-dimensional tensor-network contraction via subspace iteration}} (\bibinfo
  {year} {2026}),\ \Eprint {https://arxiv.org/abs/2607.15158} {arXiv:2607.15158
  [cond-mat.str-el]} \BibitemShut {NoStop}%
\bibitem [{\citenamefont {Chen}\ \emph {et~al.}(2026)\citenamefont {Chen},
  \citenamefont {Fu}, \citenamefont {Yu}, \citenamefont {Yu},\ and\
  \citenamefont {Xie}}]{stagger}%
  \BibitemOpen
  \bibfield  {author} {\bibinfo {author} {\bibfnamefont {H.}~\bibnamefont
  {Chen}}, \bibinfo {author} {\bibfnamefont {Y.}~\bibnamefont {Fu}}, \bibinfo
  {author} {\bibfnamefont {W.}~\bibnamefont {Yu}}, \bibinfo {author}
  {\bibfnamefont {R.}~\bibnamefont {Yu}},\ and\ \bibinfo {author}
  {\bibfnamefont {Z.~Y.}\ \bibnamefont {Xie}},\ }\bibfield  {title} {\bibinfo
  {title} {{Single-layer framework of variational tensor network states}},\
  }\href {https://doi.org/10.1103/cb6p-w59b} {\bibfield  {journal} {\bibinfo
  {journal} {Phys. Rev. B}\ }\textbf {\bibinfo {volume} {113}},\ \bibinfo
  {pages} {155127} (\bibinfo {year} {2026})}\BibitemShut {NoStop}%
\bibitem [{\citenamefont {Xie}\ \emph {et~al.}(2012)\citenamefont {Xie},
  \citenamefont {Chen}, \citenamefont {Qin}, \citenamefont {Zhu}, \citenamefont
  {Yang},\ and\ \citenamefont {Xiang}}]{PhysRevB.86.045139}%
  \BibitemOpen
  \bibfield  {author} {\bibinfo {author} {\bibfnamefont {Z.~Y.}\ \bibnamefont
  {Xie}}, \bibinfo {author} {\bibfnamefont {J.}~\bibnamefont {Chen}}, \bibinfo
  {author} {\bibfnamefont {M.~P.}\ \bibnamefont {Qin}}, \bibinfo {author}
  {\bibfnamefont {J.~W.}\ \bibnamefont {Zhu}}, \bibinfo {author} {\bibfnamefont
  {L.~P.}\ \bibnamefont {Yang}},\ and\ \bibinfo {author} {\bibfnamefont
  {T.}~\bibnamefont {Xiang}},\ }\bibfield  {title} {\bibinfo {title}
  {{Coarse-graining renormalization by higher-order singular value
  decomposition}},\ }\href {https://doi.org/10.1103/PhysRevB.86.045139}
  {\bibfield  {journal} {\bibinfo  {journal} {Phys. Rev. B}\ }\textbf {\bibinfo
  {volume} {86}},\ \bibinfo {pages} {045139} (\bibinfo {year}
  {2012})}\BibitemShut {NoStop}%
\bibitem [{\citenamefont {Menczer}\ and\ \citenamefont
  {Legeza}(2025)}]{Menczer2025}%
  \BibitemOpen
  \bibfield  {author} {\bibinfo {author} {\bibfnamefont {A.}~\bibnamefont
  {Menczer}}\ and\ \bibinfo {author} {\bibfnamefont {{\"O}.}~\bibnamefont
  {Legeza}},\ }\bibfield  {title} {\bibinfo {title} {{Massively Parallel Tensor
  Network State Algorithms on Hybrid CPU-GPU Based Architectures}},\ }\href
  {https://doi.org/10.1021/acs.jctc.4c00661} {\bibfield  {journal} {\bibinfo
  {journal} {Journal of Chemical Theory and Computation}\ }\textbf {\bibinfo
  {volume} {21}},\ \bibinfo {pages} {1572} (\bibinfo {year}
  {2025})}\BibitemShut {NoStop}%
\bibitem [{\citenamefont {Ueda}\ \emph {et~al.}(2020)\citenamefont {Ueda},
  \citenamefont {Okunishi}, \citenamefont {Yunoki},\ and\ \citenamefont
  {Nishino}}]{PhysRevE.102.032130}%
  \BibitemOpen
  \bibfield  {author} {\bibinfo {author} {\bibfnamefont {H.}~\bibnamefont
  {Ueda}}, \bibinfo {author} {\bibfnamefont {K.}~\bibnamefont {Okunishi}},
  \bibinfo {author} {\bibfnamefont {S.}~\bibnamefont {Yunoki}},\ and\ \bibinfo
  {author} {\bibfnamefont {T.}~\bibnamefont {Nishino}},\ }\bibfield  {title}
  {\bibinfo {title} {{Corner transfer matrix renormalization group analysis of
  the two-dimensional dodecahedron model}},\ }\href
  {https://doi.org/10.1103/PhysRevE.102.032130} {\bibfield  {journal} {\bibinfo
   {journal} {Phys. Rev. E}\ }\textbf {\bibinfo {volume} {102}},\ \bibinfo
  {pages} {032130} (\bibinfo {year} {2020})}\BibitemShut {NoStop}%
\bibitem [{\citenamefont {Halko}\ \emph {et~al.}(2011)\citenamefont {Halko},
  \citenamefont {Martinsson},\ and\ \citenamefont
  {Tropp}}]{Halko2011Randomized}%
  \BibitemOpen
  \bibfield  {author} {\bibinfo {author} {\bibfnamefont {N.}~\bibnamefont
  {Halko}}, \bibinfo {author} {\bibfnamefont {P.-G.}\ \bibnamefont
  {Martinsson}},\ and\ \bibinfo {author} {\bibfnamefont {J.~A.}\ \bibnamefont
  {Tropp}},\ }\bibfield  {title} {\bibinfo {title} {{Finding Structure with
  Randomness: Probabilistic Algorithms for Constructing Approximate Matrix
  Decompositions}},\ }\href {https://doi.org/10.1137/090771806} {\bibfield
  {journal} {\bibinfo  {journal} {SIAM Review}\ }\textbf {\bibinfo {volume}
  {53}},\ \bibinfo {pages} {217} (\bibinfo {year} {2011})}\BibitemShut
  {NoStop}%
\bibitem [{\citenamefont {{TeneT.jl contributors}}(nd)}]{TeneTjl}%
  \BibitemOpen
  \bibfield  {author} {\bibinfo {author} {\bibnamefont {{TeneT.jl
  contributors}}},\ }\href@noop {} {\bibinfo {title} {{TeneT.jl}}},\ \bibinfo
  {howpublished} {\url{https://github.com/XingyuZhang2018/TeneT.jl}} (\bibinfo
  {year} {n.d.}),\ \bibinfo {note} {open-source Julia package}\BibitemShut
  {NoStop}%
\bibitem [{\citenamefont {Zhang}\ \emph
  {et~al.}(2026{\natexlab{d}})\citenamefont {Zhang}, \citenamefont {Haegeman},\
  and\ \citenamefont {He}}]{ScalableTNData}%
  \BibitemOpen
  \bibfield  {author} {\bibinfo {author} {\bibfnamefont {X.-Y.}\ \bibnamefont
  {Zhang}}, \bibinfo {author} {\bibfnamefont {J.}~\bibnamefont {Haegeman}},\
  and\ \bibinfo {author} {\bibfnamefont {Y.}~\bibnamefont {He}},\ }\href@noop
  {} {\bibinfo {title} {Data and plotting scripts for ``{Unraveling square and
  kagome lattice frustrated magnetism by scalable tensor networks}''}},\
  \bibinfo {howpublished}
  {\url{https://github.com/XingyuZhang2018/scalable-tensor-network-data}}
  (\bibinfo {year} {2026}{\natexlab{d}})\BibitemShut {NoStop}%
\bibitem [{\citenamefont {Demmel}\ \emph {et~al.}(2012)\citenamefont {Demmel},
  \citenamefont {Grigori}, \citenamefont {Hoemmen},\ and\ \citenamefont
  {Langou}}]{Demmel2012CommunicationOptimalQR}%
  \BibitemOpen
  \bibfield  {author} {\bibinfo {author} {\bibfnamefont {J.}~\bibnamefont
  {Demmel}}, \bibinfo {author} {\bibfnamefont {L.}~\bibnamefont {Grigori}},
  \bibinfo {author} {\bibfnamefont {M.}~\bibnamefont {Hoemmen}},\ and\ \bibinfo
  {author} {\bibfnamefont {J.}~\bibnamefont {Langou}},\ }\bibfield  {title}
  {\bibinfo {title} {Communication-optimal parallel and sequential {QR} and
  {LU} factorizations},\ }\href {https://doi.org/10.1137/080731992} {\bibfield
  {journal} {\bibinfo  {journal} {SIAM Journal on Scientific Computing}\
  }\textbf {\bibinfo {volume} {34}},\ \bibinfo {pages} {A206} (\bibinfo {year}
  {2012})}\BibitemShut {NoStop}%
\end{thebibliography}%

\clearpage

\onecolumngrid

\crefalias{figure}{suppfigure}
\crefalias{equation}{suppequation}
\crefalias{table}{supptable}
\crefalias{section}{suppnote}
\setcounter{section}{0}
\setcounter{subsection}{0}
\setcounter{equation}{0}
\setcounter{figure}{0}
\setcounter{table}{0}
\setcounter{secnumdepth}{1}
\renewcommand{\thesection}{\arabic{section}}
\renewcommand{\theequation}{S\arabic{equation}}
\renewcommand{\thefigure}{\arabic{figure}}
\renewcommand{\thetable}{\arabic{table}}
\renewcommand{\figurename}{Supplementary Fig.}
\renewcommand{\tablename}{Supplementary Table}

\renewcommand{\theHsection}{supp.\arabic{section}}
\renewcommand{\theHequation}{supp.\arabic{equation}}
\renewcommand{\theHfigure}{supp.\arabic{figure}}
\renewcommand{\theHtable}{supp.\arabic{table}}
\makeatletter
\def\@sectioncntformat#1{Supplementary Note~\csname the#1\endcsname:\quad}
\def\@hangfrom@section#1#2#3{#1#2#3}
\def\@hangfroms@section#1#2{#1#2}
\renewcommand{\section}{\@startsection{section}{1}{\z@}
  {2.5ex \@plus 1ex \@minus .2ex}{1.2ex}
  {\normalfont\normalsize\bfseries\raggedright}}
\renewcommand{\subsection}{\@startsection{subsection}{2}{\z@}
  {2ex \@plus 1ex \@minus .2ex}{1ex}
  {\normalfont\normalsize\bfseries\raggedright}}
\def\fnum@figure{\textbf{\figurename~\thefigure}}
\def\fnum@table{\textbf{\tablename~\thetable}}
\makeatother
\pdfbookmark[0]{Supplementary Information}{supplementary-information}
\begin{center}
  {\Large\bfseries Supplementary Information}
\end{center}

\FloatBarrier
\section{Plaquette-order convergence at \texorpdfstring{$J_2/J_1=0.55$ for the $J_1$-$J_2$ square lattice model}{J2/J1 = 0.55}}
\label{app:square_vbs_j055}

\begin{figure}[htbp]
  \centering
  \includegraphics[width=0.58\textwidth]{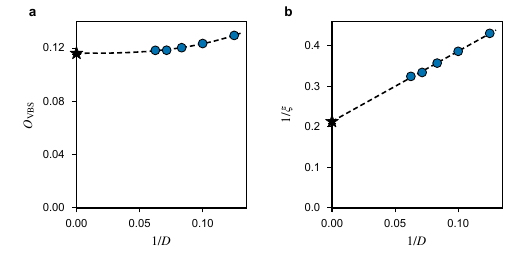}
  \caption{\textbf{\boldmath Bond-dimension $D$ convergence analysis for $J_2/J_1=0.55$ using the
  $D=8,10,12,14,16$ data.} \textbf{a} Plaquette VBS order parameter
$O_{\rm VBS}$, fitted to $O_{\rm VBS}(\infty)+a/D^{\alpha}$ with free
$\alpha$. \textbf{b} Inverse correlation length $1/\xi$ as a function of
$1/D$ with tentative linear fit. Dashed curves mark
the fits; stars mark the tentative extrapolated results. We use $\chi=512$ for all five bond dimensions.}
  \label{fig:square_j055_vbs_invD}
\end{figure}

We provide data in \cref{fig:square_j055_vbs_invD} as evidence that the plaquette order parameter converges quickly with the PEPS bond dimension $D$ for $J_2/J_1=0.55$. In the main text, we use the largest-$D$ data as a proxy for the plaquette order parameters in \cref{fig:fig1}. This convergence analysis indicates that, at least away from the boundary ($J_2/J_1=0.52$), the error of the proxy is sufficiently small.

We use the following ansatz to extrapolate the plaquette order parameter to $D \rightarrow \infty$: 
\begin{equation}
  O_{\rm VBS}(D)=O_{\rm VBS}(\infty)+\frac{a}{D^{\alpha}}.
\end{equation}
The three-parameter least-squares fit [\cref{fig:square_j055_vbs_invD}a] using all five data points gives $O_{\rm VBS}(\infty)=0.116$, and its difference from the $D=16$ data is smaller than the marker size.  To test this empirical extrapolation, we drop the two smallest-$D$ data points one by one in the fit, as a way to obtain an estimate of the extrapolation errors; the result changes by at most $3\%$. Note that we only extrapolate in $D$, as we assume that $O_{\rm VBS}(D,\chi)$ has already converged in $\chi$, with the value of $\chi$ listed in the caption.

The VBS states are believed to be gapped with finite correlation length $\xi$. This is unlike the magnetically ordered states where $\xi$ diverges as $D,\chi \rightarrow \infty$. An extrapolation of $\xi$ is in general more complicated because $\xi(D,\chi)$ is more $ \chi$-sensitive than  $O_{\rm VBS}(D,\chi)$.  As a rough inference, we plot $1/\xi$ vs. $1/D$ in~\cref{fig:square_j055_vbs_invD}b, together with a tentative linear fit.
  
\FloatBarrier
\section{Evidence for plaquette VBS from a less-restricted  $\pi$-mirror ansatz}
\label{app:plaquette_bond_symmetry}

\begin{figure}[htbp]
  \centering
  \includegraphics[width=0.48\textwidth]{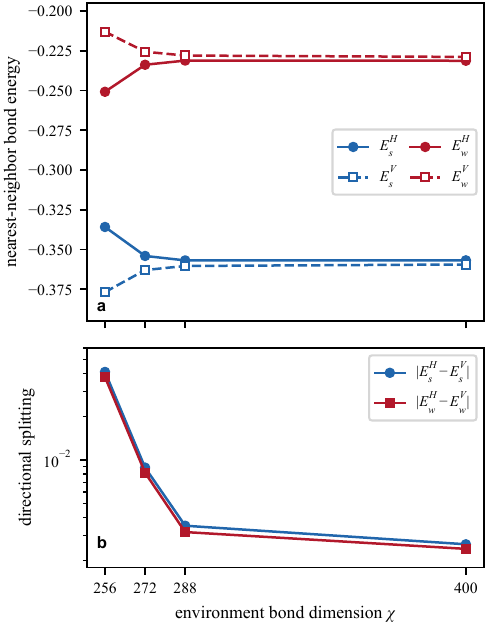}
  \caption{\textbf{\boldmath Evidence for the absence of spontaneous anisotropy for VBS at $J_2/J_1=0.55$ based on $\pi$-mirror ansatz without diagonal-reflection (isotropic) restriction.} We work with $D=10$ and show that higher-precision optimization by increasing $\chi$ can reduce the artificial anisotropy signal due to inaccuracy.  
  \textbf{a}
   The horizontal and vertical strong and weak bond energies defined in \cref{eq:less_restricted_plaquette}. \textbf{b} Directional
  splittings on a
  logarithmic scale. 
  The lines are guides to the eye.}
  \label{fig:plaquette_bond_isotropization}
\end{figure}

As introduced in the main text, our $\pi$-mirror ansatz for  the square lattice models implements two mirror symmetries in its trivial representation, thereby using one independent tensor to generate the $2 \times 2$ unit cell via horizontal and vertical flips. The ansatz is justified by its variational energy advantage. Plaquette, columnar, and anisotropic plaquette VBS orders are all compatible with this ansatz. Assuming a plaquette VBS order, we further implement the trivial representation of diagonal reflection symmetry to reduce the number of parameters and accelerate optimization. As a justification of this restriction and thus to determine whether the isotropic plaquette order is physically favored, we have also performed optimization without imposing this diagonal reflection restriction. The result~(\cref{fig:plaquette_bond_isotropization}) for $J_2/J_1=0.55$ provides confirmation for the rotationally symmetric plaquette VBS.

Our tensor plots in \cref{fig:pimirror} of the main text have implied the trivial reflection symmetries by coloring. Here we give the formal definition by tensor-index permutation in \cref{eq:less_restricted_plaquette} and \cref{eq:diagonal_reflection}. 
 We order the tensor indices as
$(1,2,3,4,5)=(l,d,r,u,p)$: left, down, right, and up virtual legs,
followed by the physical leg.  The diagram below uses the unit cell and tensor
convention of main-text \cref{fig:pimirror}a, with PEPS tensors and bond
energies labeled explicitly.  The operator ${\cal P}_{\sigma}$
permutes the tensor indices into the listed order $\sigma$.
We label the PEPS tensors as $\widetilde M_{ij}$, where $i=1,2$ runs
from top to bottom and $j=1,2$ from left to right in the $2\times2$ cell.
This less-restricted ansatz uses
\begin{equation}
 \begin{gathered}
  \includegraphics[width=0.26\textwidth]{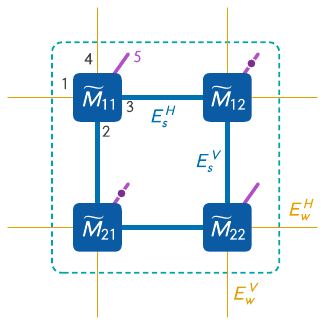}
 \end{gathered}
 \qquad
 \begin{aligned}
 \widetilde M_{11}&=M_{\rm raw},&
 \widetilde M_{12}&={\cal P}_{(3,2,1,4,5)}M_{\rm raw},\\
 \widetilde M_{21}&={\cal P}_{(1,4,3,2,5)}M_{\rm raw},&
 \widetilde M_{22}&={\cal P}_{(3,4,1,2,5)}M_{\rm raw},
 \end{aligned}
 \label{eq:less_restricted_plaquette}
\end{equation}

The trivial diagonal reflection symmetry is a restriction on a single PEPS tensor
\begin{equation}\label{eq:diagonal_reflection}
M_{\rm raw}={\cal P}_{(4,3,2,1,5)}M_{\rm raw}.
\end{equation}
Without implementing~\cref{eq:diagonal_reflection}, the horizontal
and vertical strong-bond energies, $E_s^H$ and $E_s^V$, are independent, and
likewise for the weak-bond energies, $E_w^H$ and $E_w^V$.

For $J_2/J_1=0.55$ and working at $D=10$, \cref{fig:plaquette_bond_isotropization} shows that optimization at higher precision by using larger $\chi$ points to a vanishing anisotropy, which is quantified as
\begin{equation}
 \delta_s=\left|E_s^H-E_s^V\right|,\qquad
 \delta_w=\left|E_w^H-E_w^V\right|.
 \label{eq:plaquette_directional_splitting}
\end{equation}
Increasing optimization $\chi$ from 256 to 400, $\delta_s$ decreases from $0.0409$ to
$0.0026$ while $\delta_w$ decreases from $0.0377$ to $0.0024$. We consider the small residual as a finite $\chi$ and finite $D$ effect remaining. 

Additional justification comes from the comparison with data with diagonal mirror symmetry restriction. For $D=10,\chi=400$, the optimized energy of unrestricted and restricted ansatz are
$-0.486225$ and $-0.486213$ per site respectively.  The energy difference ($10^{-5}$) is small compared to their difference to the ground state energy ($>1\times 10^{-4}$, bounded by our $D=16$ result).  

\FloatBarrier
\section{VBS susceptibilities from finite-correlation-length scaling }
\label{app:vbs_scaling}
\begin{figure}[htbp]
  \centering
  \includegraphics[width=0.45\textwidth]{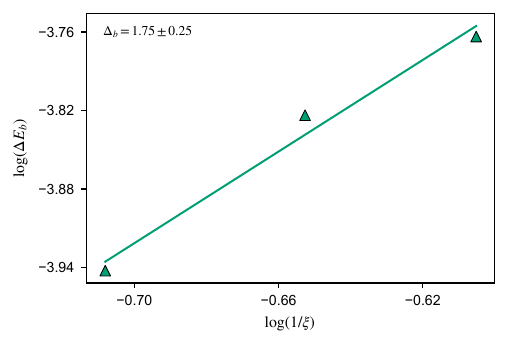}
  \caption{\textbf{\boldmath Finite-correlation-length log--log fit of the kagome $2\times2$ absolute
  bond-energy range $\Delta E_b=\max E_b-\min E_b$ over all 24 bonds.} The
  three $D=8,9,10$ points use $\chi=1024$ and the physical lattice-spacing
  correlation length. Assuming $\Delta E_b \propto 1/\xi^{\Delta_b}$, the fit gives $\Delta_b=1.75(25)$.}
  \label{fig:kagome_bond_range_scaling}
\end{figure}

In the main text, we have performed finite-correlation-length scaling for the antiferromagnetic susceptibilities of the kagome Heisenberg model based on the $1\times 1$ unit-cell data, shown in \cref{fig:square_kagome_m2_xi}. As we show in the inset of \cref{fig:kagome_energy_xi3} in the main text, the $2\times 2$ unit-cell data have a dominant finite-$D$ VBS order. For a gapless spin liquid, this order is also supposed to scale to zero with increasing $\xi(D,\chi)$. We characterize the finite-$D$ VBS order as $\Delta E_b=\max (E_b)-\min(E_b)$, where the maximum and minimum are taken over all 24 bonds. We plot $\Delta E_b$ vs. $1/\xi$ in \cref{fig:kagome_bond_range_scaling}. For a gapless spin liquid, we expect finite-correlation-length scaling $\Delta E_b \propto 1/\xi^{\Delta_b}$, where $\Delta_b > 1.5$ is again required in order for the VBS susceptibility to be stable against forming long-range order. The effective theory~\cite{PhysRevB.77.224413,Song2019} of the  U(1) Dirac spin liquid predicts emergent symmetry leading to $\Delta_b=\Delta_m$, where $\Delta_m$ is the spin scaling dimension of \cref{fcs}. The fit gives $\Delta_b=1.75(25)$, slightly below but consistent within the error bars with $\Delta_m=1.97(6)$ from the main text. Data at larger $D$ would sharpen this comparison and the margin above $\Delta_b=1.5$.

\FloatBarrier
\section{Optimization and evaluation of iPEPS at different environment bond dimensions lead to deviations in extrapolations}
\label{app:chi_reoptimization}

\begin{figure}[htbp]
  \centering
  \includegraphics[width=0.45\textwidth]{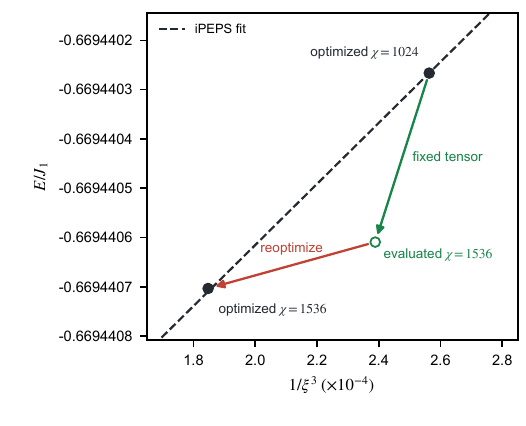}
  \caption{\textbf{\boldmath Illustration of evaluation at larger $\chi$ and subsequent
  reoptimization, using $D=16$ Heisenberg data.}
  Filled circles are optimized and evaluated at $\chi=1024$ and $1536$;
  the open circle evaluates the unchanged $\chi=1024$ tensor at $\chi=1536$.
  The green arrow increases only the evaluation dimension; the red arrow
  follows 50 optimization steps at the larger $\chi$. The dashed line reproduces
  the original iPEPS energy fit from main-text \cref{fig:squareheisenberg}a,
  without fitting the three points shown here.}
  \label{fig:chi_reoptimization}
\end{figure}

We illustrate the distinction between increasing the environment dimension
used to evaluate an iPEPS and reoptimizing its tensor at the larger dimension.
\cref{fig:chi_reoptimization} uses numerical data for a
$C_{4v}$ Heisenberg iPEPS with $D=16$ as an illustrative example.
A tensor optimized at $\chi=1024$ is first evaluated at the same $\chi$,
then reevaluated at $\chi=1536$ without changing its entries. It is
subsequently used to initialize optimization at $\chi=1536$.

Increasing only the evaluation dimension lowers the energy by
$3.43\times10^{-7}J_1$, while $\xi$ grows from $15.7420$ to $16.1171$
(only $2.4\%$, green arrow). Starting from this same tensor, just 50
L-BFGS optimization steps at $\chi=1536$ lower the energy by a further
$9.41\times10^{-8}J_1$, but increase $\xi$ to $17.5583$
($8.9\%$, red arrow). Thus, increasing $\chi$ mainly improves the energy
estimate in this example, whereas a short optimization continuation yields
a much larger increase in correlation length despite its smaller energy gain.
The $D=12,14$ fit from main-text \cref{fig:squareheisenberg}a is shown as an
independent reference: both optimized points lie close to it, whereas
the fixed-tensor reevaluation departs from it. The comparison illustrates
why increasing $\chi$ only during evaluation does not generally reproduce
the sequence of tensors optimized and evaluated at the same $\chi$
used in the one-shot scaling analysis.

\FloatBarrier
\section{Randomized loop contraction for evaluating  diagonal \texorpdfstring{$J_2$}{J2} coupling}
\label{app:j2_projector}

In Methods, we have introduced how to use a randomized low-rank method to reduce the computational complexity of computing the diagonal spin-exchange coupling $J_2\sum_{\langle\langle ij \rangle\rangle}\mathbf{S}_i\cdot\mathbf{S}_j$ in the Hamiltonian of the $J_1$-$J_2$ square lattice model. Here we provide more details of the method and the numerical evidence that the precision is maintained for practical purposes.

\begin{figure}[htbp]
  \centering
  \includegraphics[width=0.30\columnwidth]{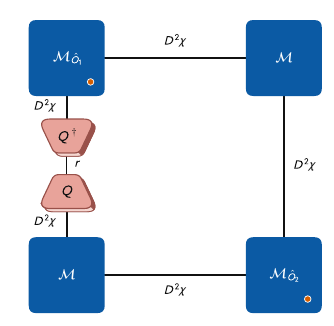}
  \caption{\textbf{Randomized loop contraction.} Each blue contracted corner
  tensor $\mathcal M$ is an $N\times N$ matrix with $N=D^2\chi$. The isometry
  $Q_{\hat O, \hat O}:\mathbb{C}^{\bar{\chi}}\rightarrow\mathbb{C}^{N}$ and its adjoint replace one
  full bond by a rank-$\bar{\chi}$ channel. $Q_{\hat O, \hat O}$ is built by \cref{eq:j2_range_finder} (\cref{eq:qiter}).}
  \label{fig:j2_projector}
\end{figure}

As a reiteration, the goal is to perform the contraction illustrated in \cref{fig:pimirror}b of the main text. The idea is to use a randomized low-rank method to find the projected basis $Q$ [\cref{fig:pimirror}d] that captures the approximate low-rank structure to reduce the cost of contraction. We replicate \cref{fig:pimirror}d from the main text to \cref{fig:j2_projector}.  
 After contracting the physical indices and grouping each
corner with its adjacent boundary tensors, every corner becomes a matrix $\mathcal M$
whose two fused indices have dimension $D^2\chi$. The operator-bearing corners are both denoted by $\mathcal M_{\hat O}$. Cutting one bond of the resulting four-matrix loop defines an
$N\times N$ effective transfer operator
\begin{equation}
  K_{\hat O,\hat O} = \mathcal M_{\hat O} \mathcal M \mathcal M_{\hat O} \mathcal M,
  \qquad Z_{\hat O,\hat O} = \operatorname{Tr} K_{\hat O,\hat O} .
  \label{eq:j2_direct}
\end{equation}

For next-nearest neighbor $\langle \langle ij\rangle \rangle$, the expectation value of the spin-exchange coupling on the variational state is calculated as 
\begin{equation} \label{eq:J2energy}
\frac{\langle \psi |\mathbf{S}_i\!\cdot\!\mathbf{S}_j| \psi \rangle} {\langle \psi | \psi \rangle} = \frac{ Z_{\hat S_x,\hat S_x} + Z_{\hat S_y,\hat S_y} + Z_{\hat S_z,\hat S_z} }{ Z_{\hat I,\hat I} } \end{equation}
Forming $K_{\hat O,\hat O}$  has computational complexity  $ O(D^6\chi^3)$, which is higher than the VUMPS iteration complexity.

We implement the randomized QB method~\cite{Halko2011Randomized} to approximately evaluate $Z_{\hat O,\hat O}$ without forming $K_{\hat O,\hat O}$. The randomized QB method is known to be able to recover exact low-rank structure and is good at revealing approximate low-rank structure. Let
$\bar{\chi}=\chi+p$ and $\Omega\in\mathbb{C}^{N\times \bar{\chi}}$ be a Gaussian random matrix. With the assumption that $\bar{\chi} = \chi$ may already be a good low-rank approximation for the projector to be constructed, the choice of $p>0$ is dubbed oversampling.  We calculate a matrix $Q_{_{\hat O,\hat O}}$ representing the projected basis satisfying $Q_{_{\hat O,\hat O}}^\dagger Q_{_{\hat O,\hat O}}=I$, where $I$ is the identity matrix. It can be obtained from 
\begin{equation}
  Y_{\hat O,\hat O} = K_{\hat O,\hat O} \Omega, \qquad Y_{\hat O,\hat O} = Q_{\hat O,\hat O}R, 
  \label{eq:j2_range_finder}
\end{equation}
where the second equation is a QR decomposition.
With $Q$s, we can do the approximate loop contraction
\begin{equation}
  Z_{\hat O,\hat O}
  \approx \operatorname{Tr}\!\left(Q_{\hat O,\hat O}^\dagger K_{\hat O,\hat O}Q_{\hat O,\hat O}\right),
  \label{eq:j2_projected_value}
\end{equation}
as illustrated in
\cref{fig:j2_projector}.  The above is a repetition of the minimal implementation of randomized loop contraction we introduce in Methods.  In the randomized QB method, more subspace iterations $q$ to construct the $Y$ matrix may help improve the contraction precision
\begin{equation}\label{eq:qiter}
 Y^{(q)}_{\hat O,\hat O}=(KK^\dagger)^qK\Omega
\end{equation}

Computing $Y$ costs
$O(D^4\chi^2\bar{\chi})$, while the  QR factorization costs $O(D^2\chi\bar{\chi}^2)$.
Under the assumption that $\bar{\chi}=O(\chi)$ is sufficient to keep high precision, this cost amounts to
\begin{equation}
   O(D^4\chi^3),
  \label{eq:j2_projected_cost}
\end{equation}
which is of the same order as the VUMPS iteration complexity.

\begin{figure}[htbp]
  \centering
  \includegraphics[width=0.67\textwidth]{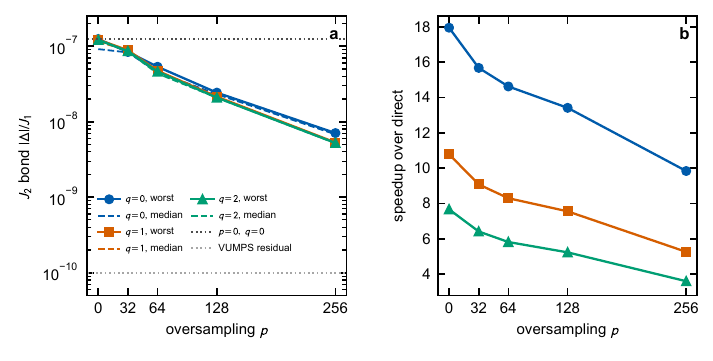}
  \caption{\textbf{Evidence for the high precision of randomized loop contraction for diagonal spin-exchange coupling energy and its speed-up over direct contraction.} The effects of algorithm parameters $q$ and $p$ are investigated. We contract  $D=8$ iPEPS optimized with
  $\chi=128$ for the $J_1$-$J_2$ square lattice model at  $J_2/J_1=0.5$. We use three random seeds in the contraction. \textbf{a} Worst-seed (solid) and median (dashed) $J_2$ coupling energy error in units of $J_1$. Two dotted horizontal lines denote the minimal-cost $p=0$, $q=0$ reference and the VUMPS boundary residual, respectively.
  \textbf{b} Speedup of randomized contraction over direct contraction. The test is performed using an H100 GPU which takes  $6.859\,\mathrm{s}$ for direct contraction. The color scheme for $q$ follows a.}
  \label{fig:j2_pq_scan}
\end{figure}

Now we provide evidence that $\bar{\chi}=O(\chi)$ is sufficient in our calculation, and benchmark the choice of $\bar{\chi}=\chi+p$ and iteration $q$. For relatively small $D$, we can compute~\cref{eq:J2energy} without the randomized low-rank method. Thus, we can validate \cref{eq:j2_projected_value} against
\cref{eq:j2_direct}. As an example, we work with $D=8$, $J_2/J_1=0.5$.  For an optimized iPEPS and well-converged VUMPS environment (residuals below $10^{-10}$), we compute the diagonal coupling energies.  The results for $\chi=128$ are summarized in \cref{fig:j2_pq_scan}. We use three random seeds to construct $\Omega$. \cref{fig:j2_pq_scan}a indicates that the fluctuation of precision is small over random seeds. The errors here are around $10^{-7}$ and can be efficiently reduced by increasing $p$. As a reference, we work with Float64 precision ($10^{-15}$-$10^{-14}$), and the VUMPS convergence residual is about $10^{-10}$. A precision of $10^{-7}$ is sufficient for our calculation of the $J_1$-$J_2$ model where our best ground-state energy precision is estimated as $10^{-5}$. 
The results of \cref{fig:j2_pq_scan} indicate that, in order to improve precision, using more iterations, i.e.\ $q>0$, is less effective than oversampling $p$. We also benchmark $\chi=256,512$ for $D=8$. The calculation of $q=0, p=0$ gives errors 
$1.48\times10^{-8}$, and $5.52\times10^{-8}$ averaged over three fixed random seeds, compared to $1.73\times10^{-7}$ for $\chi=128$.

\begin{figure}[htbp]
  \centering
  \includegraphics[width=0.58\textwidth]{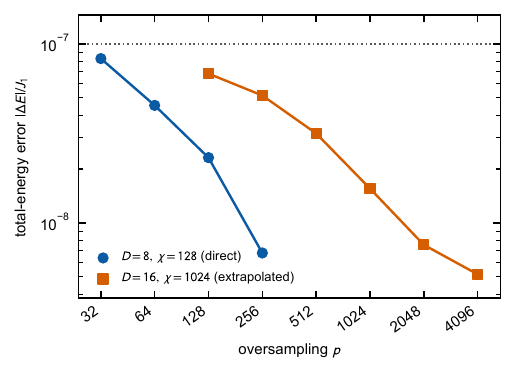}
  \caption{\textbf{\boldmath Randomized loop contraction for $D=16$ compared with $D=8$~(\cref{fig:j2_pq_scan}).}
  The $D=16$ iPEPS is also optimized for $J_2/J_1=0.5$ and with $\chi=1024$. We set the iteration parameter $q=0$ and investigate the precision control by oversampling $p$. The results shown are the three-seed median. As a proxy for the expensive direct contraction calculation for $D=16$, we use polynomial $1/p$ extrapolation up to the quadratic term as a reference.}
  \label{fig:j2_projector_error_p}
\end{figure}

We also benchmark randomized contraction in larger $D$, for which direct contraction \cref{eq:j2_direct} is not available as a reference within currently reasonable computational resources. As an example, we contract the $D=16$ state optimized with $\chi=1024$.  
Direct contraction is not feasible at $D=16$, so we use large oversampling $p$ extrapolation as the proxy $E(D=16,\chi=1024,p=\infty)=-0.496704261686$.
The estimation errors and speed-up factor as a function of oversampling $p$ are shown in~\cref{fig:j2_projector_error_p}; the $D=8$ data are replicated for comparison. We see that for the same ratio of $p/\chi$, the errors in $D=16$ are reduced.

\FloatBarrier
\section{Parallel algorithms for differentiable VUMPS}
\label{app:multigpu}
In Methods, we have explained a tensor contraction slicing scheme avoiding the naive intermediate large  memory peak, which also naturally leads to a simple efficient parallelization scheme (1DSlice).  The key idea is to divide the data of tensors by slicing one index with dimension $\chi$ to several parts.  Evaluating each part in series  leads to reduction of peak memory.   Evaluating parts in parallel gives the 1DSlice parallelization scheme. In this note, we first supplement the details of the technique to realize the $O(D^2\chi^2)$-per-device memory cost (1DSlice) parallel optimization algorithm. We next explain our 2DSlice parallelization scheme, which extends the 1DSlice scheme and distributes the $O(D^2\chi^2)$ memory to devices, lifting the limitation set by the single device memory. Finally, we show the profiling results for implementation of our methods on GPU devices across different computers. The results support that our parallelization scheme is near-ideal.

\subsection{Slicing dominant environment tensor maps }
Here, we extend the discussion of the slicing scheme introduced in Methods. 

\begin{figure}[htbp]
\centering
\includegraphics[width=0.985\textwidth]{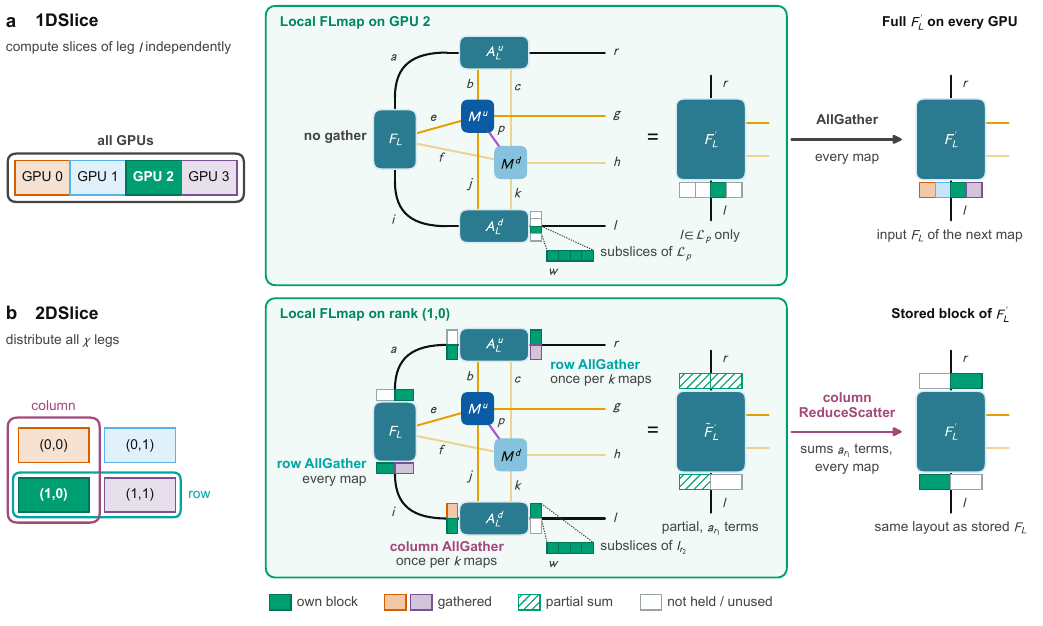}
\caption{\textbf{1DSlice and 2DSlice parallelization schemes for FLmaps (4 GPUs for example).}
Device (GPU) organization as 1D and 2D grids of MPI ranks (left); Data distribution and task for FLmap on one GPU, GPU 2 for 1DSlice and GPU (1,0) for 2DSlice as example  (center); Collective communication concluding one FLmap (right). Color scheme denotes the one-to-one correspondence between data block (index slices) and GPU. For example, green cells denote the blocks owned by GPU 2 and GPU (1,0) (persistently stored and/or computed). The GPU receives data from other GPUs through an AllGather with the corresponding colors. White cells denote the block never held or used in the GPU. For 2DSlice, striped cells are partial sums.
\textbf{a} In 1DSlice, every GPU stores the
complete $F_L$, $A_L^u$, $A_L^d$, and $M$ (both layers $M^u$ and $M^d$). GPU $p$ evaluates  $(F_L')_{\mathcal L_p}$ where $\mathcal{L}_p$ (green block) is a slice of index $l$. To reduce peak memory of each GPU, subslices of $(F_L')_{\mathcal L_p}$ are computed one by one, with width $w$. An AllGather
over all GPUs collects the full $F_L'$ on every GPU, and we use $F_L \leftarrow F_L'$ in the next map.
\textbf{b} In 2DSlice on an $N_1\times N_2$ grid (shown $2\times2$), 
GPU rank $(r_1,r_2)$
persistently stores the blocks $F_L[a_{r_1},e,f,i_{r_2}]$, $A_L^u[a_{r_1},b,c,r_{r_2}]$,
$A_L^d[i_{r_1},j,k,l_{r_2}]$, and $F_L'[r_{r_1},g,h,l_{r_2}]$ each with size $S_E/P$. $M$ is complete on every rank. The operations are shown  for GPU rank
$(r_1,r_2)=(1,0)$. Row AllGathers
provide the full $i$ leg of $F_L$ and the full $r$ leg of $A_L^u$, and a column AllGather provides the full $i$ leg of
$A_L^d$. The gathered data sizes are   $S_E/N_1$ and $S_E/N_2$ respectively. The $F_L$ slices are gathered in
every map application. As we do $k$ iterations, the $A_L$ slices are gathered once per $k$
applications because $A_L$ is fixed during the iteration. The local contraction produces the partial
output $\tilde F_L'[r,g,h,l_{r_2}]$, complete in $r$ but containing only the
$a_{r_1}$ terms of the sum over $a$. A column ReduceScatter sums these terms
and scatters  $F_L'[r_{r_1},g,h,l_{r_2}]$ to GPU $(r_1, r_2)$, which has layout of the
input $F_L$. For the next iteration, we use $F_L[r_{r_1},g,h,l_{r_2}] \leftarrow F_L'[r_{r_1},g,h,l_{r_2}] $.}
\label{fig:parallel_slicing_schematic}
\end{figure}

Most of the computational cost in the VUMPS fixed-point environment solve~\cite{PhysRevB.97.045145,Nietner2020efficient} comes from repeated applications of the left, right, and center maps,
\begin{align}
  F_L'&={\cal F}_L(F_L,A_L^u,A_L^d,M),\nonumber\\
  F_R'&={\cal F}_R(F_R,A_R^u,A_R^d,M),\nonumber\\
  A_C'&={\cal F}_C(A_C,F_L,F_R,M).
  \label{eq:vumps_maps}
\end{align}
Each of environment tensors ($F_L$,$F_R$,$A_L^u$,$A_L^d$) has dimensions
$\chi\times D\times D\times\chi$, while the PEPS tensor $M$ has dimensions $D \times D\times D\times D \times d$. In practice, $\chi=O(D^2)$, and $d$ is fixed by the Hamiltonian. These maps [Eq.~\eqref{eq:vumps_maps}] are contractions of input tensors. Except for the input data of tensors representing different environment tensors, all maps are identical. Thus we can focus on the first map, which we dub  FLmap. The contraction is illustrated in \cref{fig:flmap_contraction} of the main text method section, with the same index labels as in \cref{fig:parallel_slicing_schematic}, and  can be written as
\begin{equation}
  (F_L')_{rghl}=\sum_{ i j e k p f a b c}
  (F_L)_{aefi}(A_L^d)_{ijkl}
  M^u_{ejgbp}M^d_{fkhcp}(A_L^u)_{abcr}.
  \label{eq:flmap_indices}
\end{equation}

To perform the contraction with minimal computational cost, the order of summation (contraction) should follow the summation order in Eq.~\eqref{eq:flmap_indices}, i.e.,  $ijekpfabc$  from left to right, in other words, contracting the tensors one by one from left to right as mentioned in the caption of \cref{fig:flmap_contraction}. The computational cost is
\begin{align}
O(\chi^3D^4)+O(\chi^2D^6d).
\end{align} 
\noindent\begin{minipage}{\linewidth}
The first intermediate,
$\sum_i(F_L)_{aefi}(A_L^d)_{ijkl}$, carries indices $aefjkl$ and contains
$\chi^2D^4$ entries. Further contraction with $M^u$ yields a second intermediate with indices $agfbpkl$ and thus $\chi^2 D^4 d$ entries. Because $l$ remains open throughout this procedure, each slice of $l$ of $F'_L$
can be evaluated independently. We define the width
$w$ as the number of consecutive $l$ components contracted together
on one device. The intermediate storage of the contraction
then gets reduced to
\begin{align}\label{eq:temporaray_memory}
O(w\chi D^4d),
\end{align}
compared with the naive $O(\chi^2 D^4d)$ peak. The case $w=1$ evaluates one
component of $l$ at a time; larger widths, such as $w=4$ or $8$, can improve
GPU utilization.  
\end{minipage}
\par

For gradient evaluation, automatic differentiation needs the forward
intermediates of \cref{eq:flmap_indices}. We recompute them one width-$w$ slice at a time and release them after applying its vector-Jacobian
product (VJP). This preserves the reduced scaling of temporary memory given by~\cref{eq:temporaray_memory}.

\subsection {Parallel computing on devices with separate memory (1DSlice)}

\begin{figure}[htbp]
\centering
\includegraphics[width=0.91\textwidth]{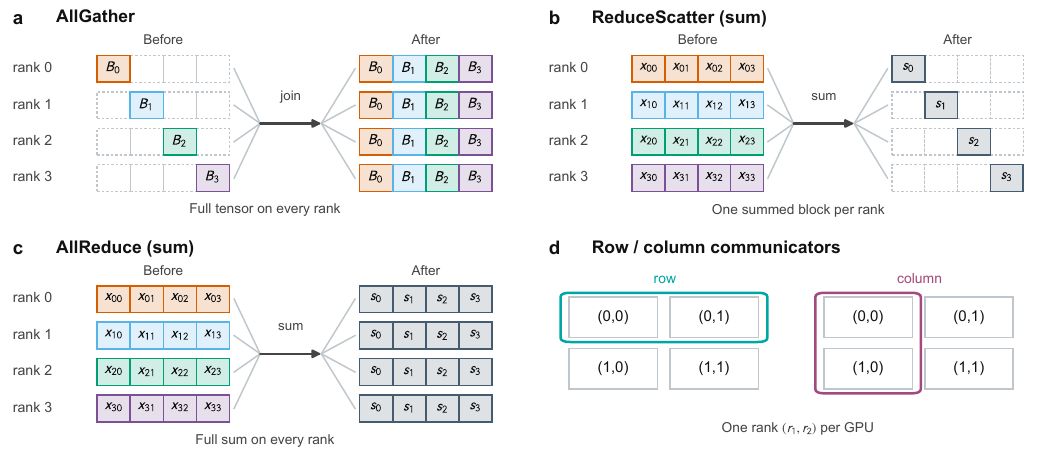}
\caption{\textbf{Collective communication operations of MPI (4 GPUs for example).} This is a brief reference to standard MPI communication operations we use. We use one process, each of which is assigned an MPI rank, per GPU. Each row is one rank, columns index blocks, dashed slots are
blocks a rank does not hold, and colors identify input ranks.
\textbf{a} AllGather concatenates disjoint blocks $B_p$ and returns the
assembled tensor to every rank. \textbf{b} ReduceScatter computes
$s_j=\sum_{p=0}^{3}x_{pj}$ and retains only the assigned block $s_p$ on
rank $p$. \textbf{c} AllReduce returns the complete sum to every rank; it is
equivalent to the ReduceScatter in \textbf{b} followed by the AllGather in
\textbf{a}. \textbf{d} Row and column communicators
on a $2\times2$ GPU grid.}
\label{fig:collective_communication}
\end{figure}

As introduced in Methods, 1DSlice evaluates $l$-index slices of
$F'_L$ in parallel on devices (GPUs) with separate memory. Let $\mathcal L_p$ denote the slices of $l$ index assigned
to MPI (Message Passing Interface, a standard message-passing system) rank $p$, where $p$ is also a label of an individual GPU. Since GPUs are usually identical,  $\mathcal L_p$ is usually an equal partition
across $P$ devices, or as equal as possible. The GPU $p$ computes  $(F_L')_{\mathcal L_p}$ slices of $F_L'$.  As long as $P \ll \chi$, to avoid possibly large intermediate peak memory explained above, each GPU evaluates a width  $w$ subslice of $(F_L')_{\mathcal L_p}$ at a time. After this local computation, each
GPU holds only its assigned part of $F'_L$.  However, the
complete updated environment on every GPU is needed for the next iteration; this is accomplished via a communication operation known as AllGather in MPI. The MPI collective communication operations we use are illustrated in \cref{fig:collective_communication}a--c; each acts within a communicator, the group of ranks that take part in it [\cref{fig:collective_communication}d].

We count the communication data sent per
GPU during one tensor map. Let
\begin{equation}
S_E=q\chi^2D^2,\qquad S_M=qdD^4
\label{eq:tensor_bytes}
\end{equation}
be the sizes in bytes of one environment tensor and the site tensor $M$,
with $q$ bytes per scalar. 

The AllGather at the end of an FLmap has the communication data volume 
\begin{equation}
V_{\rm 1D}^{\rm fwd}\simeq c_{\rm AG,E}S_E,
\label{eq:slice1d_comm_fwd}
\end{equation}
where $c_{\rm AG,E}$ denotes the coefficient that depends on the implementation details of AllGather.

In the backward propagation of FLmap to compute the gradient, each GPU propagates the adjoints through
its own part of forward FLmap.  Adjoints  are the derivatives of  energy with respect to the corresponding tensors, denoted by bars. The local VJP is
\[
\left(\overline F_L^{[p]},\overline A_L^{u,[p]},
\overline A_L^{d,[p]},\overline M^{[p]}\right)
=
\left[\mathrm D{\cal F}_L\!\left(
F_L,A_L^u,(A_L^d)_{\mathcal L_p},M\right)\right]^*
(\overline{F_L'})_{\mathcal L_p}.
\]
The superscript $[p]$ marks the local contribution before communication.
Here $\mathrm D{\cal F}_L$ is the Jacobian of the dependent variables, and
its adjoint action propagates the output adjoint to the four inputs by
the chain rule. To reduce peak memory, we also subslice $\mathcal L_p$ and evaluate $w$ components at a time. The shared-input contributions are summed locally, while
$\overline A_L^{d,[p]}$ contains disjoint $l$ intervals. The complete
input adjoints are
\[
\begin{aligned}
\overline F_L&=\operatorname{AllReduce}(\overline F_L^{[p]}),&
\overline A_L^u&=\operatorname{AllReduce}(\overline A_L^{u,[p]}),\\
\overline A_L^d&=\operatorname{AllGather}(\overline A_L^{d,[p]}),&
\overline M&=\operatorname{AllReduce}(\overline M^{[p]}).
\end{aligned}
\]
Here $\overline M^{[p]}$ includes the contributions through both PEPS
layers. Replicated copies represent one logical tensor, so each output
interval contributes to the VJP only once. The communication volume is
therefore
\begin{equation}
V_{\rm 1D}^{\rm VJP}\simeq
(2c_{\rm AR,E}+c_{\rm AG,E})S_E+c_{\rm AR,M}S_M.
\label{eq:slice1d_comm_vjp}
\end{equation}
The last term communicates only the PEPS tensor gradient. Its size ratio
$S_M/S_E=dD^2/\chi^2$ is at most $0.11\%$ for the benchmarks below,
so it constitutes only a small contribution to the volume. For equal partitions with $w\leq\chi/P$, the
per-GPU working memory in bytes is
\begin{equation}
\mathrm{Mem}_{\rm 1D}
=O\!\left(S_E+q\chi wD^4d\right).
\label{eq:slice1d_memory}
\end{equation}
The intermediate temporary memory depends on $w$ and is independent of $P$. Each GPU has persistent memory of complete environment tensors.

\subsection{Parallel computing with distributed memory (2DSlice)}

As we discussed, the 1DSlice parallel scheme has small communication cost, and the number of devices (GPUs) $P$ can scale with $\chi$. Since every device in 1DSlice parallelization has a full copy of environment tensors with memory $S_E=O(D^2\chi^2)$ and other data with similar sizes, the memory is not distributed.  To overcome the limitation that data may not fit in a single device's memory, we design a distributed memory parallelization scheme. Two indices of each environment tensor are sliced in this scheme, thus dubbed 2DSlice. This scheme has even smaller communication data, decreasing with the number of devices as $1/\sqrt{P}$. The number of communication operations, related to latency time cost, is another important quantity for practically efficient parallelization.  A design principle of 2DSlice is striking a balance between memory saving and the number of communication operations.

We arrange the devices (e.g., GPUs) as an
$N_1\times N_2$ grid, $P=N_1N_2$, as shown in
\cref{fig:parallel_slicing_schematic}b. Row index $r_1=0,1,\dots,N_1-1$ and column index $r_2=0,1,\dots,N_2-1$ are assigned to devices as well as blocks of environment tensors.
The slicing indices of environment tensors are the two bonds with dimension $\chi$.

In 2DSlice, each device persistently stores its assigned blocks of the environment tensors.  Each device never needs the full environment tensor in evaluating environment maps such as \cref{eq:flmap_indices}. Thus the communication data is reduced.  Now we focus on FLmap.  The distributed input and output tensors are
\begin{equation}
  \begin{gathered}
  F_L[a_{r_1},e,f,i_{r_2}],\qquad
  A_L^u[a_{r_1},b,c,r_{r_2}],\\
  A_L^d[i_{r_1},j,k,l_{r_2}],\qquad
  F_L'[r_{r_1},g,h,l_{r_2}].
  \end{gathered}
\end{equation}
 
Similar to the single-device and 1DSlice parallel contraction,  to reduce the naive intermediate large memory peak, each device divides its local $l_{r_2}$ range into
small slices containing at most $w$ consecutive values. Let $\alpha_{r_2}$
be one such slice. Devices in column $r_2$
compute different $a_{r_1}$ contributions for the same slice, while
different columns process different $l$ blocks. The first contraction is
\begin{equation}\label{eq:eecontract}
\left.\sum_i(F_L)_{aefi}(A_L^d)_{ijkl}
\right|_{a\in a_{r_1},\,l\in\alpha_{r_2}}.
\end{equation}
Rank $(r_1,r_2)$ stores $F_L[a_{r_1},e,f,i_{r_2}]$ and
$A_L^d[i_{r_1},j,k,l_{r_2}]$, but needs the full range of $i$ for contraction. A row
AllGather of $F_L$ and a column AllGather of $A_L^d$ provide data for the full range
[\cref{fig:collective_communication}]. We call the resulting tensor blocks
row or column gathered slices: full range of one index is gathered,
while the other remains distributed. The row and column gathered slices contain $S_E/N_1$ and $S_E/N_2$ bytes,
respectively, compared with persistent storage $S_E/P$ per environment
tensor.

For the last contraction with $A_L^u$ in the FLmap (\cref{eq:flmap_indices}), a row AllGather is first performed to provide full range of $r$ index for $A_L^u$. Since the result contains contributions from all slices of the $a$ index, a column ReduceScatter is performed over slices of $a$, yielding 
$F_L'[r_{r_1},g,h,l\in\alpha_{r_2}]$.

An alternative strategy exchanges blocks one at a time, as in Cannon's
parallel matrix multiplication, to avoid storing full gathered slices.
In practice, we find that gathering the row and column slices at once
can have much lower latency.

In our VUMPS implementation, we usually perform  $k$ FLmap iterations at a time. After the first time of row- and column AllGather for distributed $A^u_L$ and $A^d_L$,  we can keep them in memory during the $k$ iterations.
Taking into account the above communication processes,  the size of communication data per FLmap iteration (forward part of AD calculations) is estimated as
\begin{align}
\overline V_{\rm 2D}^{\rm fwd}
&\simeq S_E\left[
\frac{c_{\rm AG}^{\rm row}}{N_1}
+\frac{c_{\rm RS}^{\rm col}}{N_2}
+\frac{1}{k}\left(
\frac{\widetilde c_{\rm AG}^{\rm row}}{N_1}
+\frac{\widetilde c_{\rm AG}^{\rm col}}{N_2}\right)\right].
\label{eq:slice2d_comm_fwd}
\end{align}

Following the same VJP analysis as in \cref{eq:slice1d_comm_vjp}, with
fixed-input gathers and gradient reductions performed once per solve,
the mean volume over $k$ map VJPs is
\begin{align}
  \overline V_{\rm 2D}^{\rm VJP}
  &\simeq S_E\left[
  \frac{c_{\rm AG}^{\rm row}+c_{\rm RS}^{\rm row}}{N_1}
  +\frac{c_{\rm AG}^{\rm col}}{N_2}\right.\nonumber\\
  &\quad\left.{}+\frac{1}{k}\left(
  \frac{\widetilde c_{\rm AG}^{\rm row}+\widetilde c_{\rm RS}^{\rm row}}{N_1}
  +\frac{\widetilde c_{\rm AG}^{\rm col}+\widetilde c_{\rm RS}^{\rm col}}{N_2}
  \right)\right]+c_{\rm AR,M}S_M.
  \label{eq:slice2d_comm_vjp}
\end{align}
The tilded coefficients are evaluated for the once-per-solve messages and
need not equal those for the per-map slice messages. If the fixed $A_L$
gathered slices are retained from the forward solve, their tilded AllGather terms vanish.
For a balanced grid and implementations with bounded volume coefficients,
these environment terms scale as $O(S_E/\sqrt{P})$, compared with $O(S_E)$
in 1DSlice. The remaining global AllReduce communicates only the gradient
of $M$. Relative to the gathered-slice scale,
$S_M/(S_E/\sqrt P)=\sqrt P\,dD^2/\chi^2$, below $0.44\%$ throughout the
map benchmarks, including $P=16$.

The row and column gathered slices are transient. For equal blocks with
$w\leq\chi/N_2$, each local FLmap intermediate retains $\chi/N_1$ values
of $a$ and at most $w$ values of $l$. The per-GPU working-memory bound is
\begin{equation}
  \mathrm{Mem}_{\rm 2D}=O\!\left(
  \frac{S_E}{P}+\frac{S_E}{N_1}+\frac{S_E}{N_2}
  +\frac{q\chi wD^4d}{N_1}\right).
  \label{eq:slice2d_memory}
\end{equation}
For fixed $w$, the intermediate term decreases as $1/N_1$.
On a balanced grid, both this term and the gathered slices decrease as
$1/\sqrt P$, while the persistent environment decreases as $1/P$.
These estimates describe the dominant storage of a map and its VJP;
replicated site tensors and other solver state must also be included in
the full memory budget. The generic MPI path
supports nondivisible $\chi$ by quotient-and-remainder partitioning,
provided every block is nonempty. The NCCL path used in the benchmarks
requires equal blocks, $\chi\bmod N_1=\chi\bmod N_2=0$.

Besides the dominant environment maps, VUMPS has a canonicalization step
with subdominant computational cost $O(D^2\chi^3)$ that relates $A_R$, $A_L$, and $A_C$.
An algorithm with distributed memory is needed to keep the environment tensor storage distributed. Our implementation uses distributed tall-skinny
QR (TSQR) decomposition~\cite{Demmel2012CommunicationOptimalQR}. Additional data communications are needed for the preparation for TSQR.  A row
AllGather first assembles a $(\chi D^2/N_1)\times\chi$ input panel for
local QR. A column AllGather then assembles the $N_1$ local
$\chi\times\chi$ triangular factors for a second QR; each rank retains
the required isometry block. The initial panel contains $S_E/N_1$ data,
while the stacked factors contain $N_1q\chi^2$ bytes and contribute a
separate communication volume $c_{\rm AG}^{\rm QR}N_1q\chi^2$ per rank.
The two stages require $O(\chi^3D^2/N_1+N_1\chi^3)$ work per rank.
The right-gauge LQ construction (TSLQ) follows the transposed procedure,
with $N_2$ replacing $N_1$ and row and column collectives interchanged.
Reverse differentiation recomputes these gathers, applies the corresponding
ReduceScatters to panel and factor adjoints, and AllReduces the adjoint
of the replicated gauge matrix $C$. At fixed process grid, the QR/LQ
work is subleading in $D$, and factor communication and storage lack the
$D^2$ factor of the gathered slices. The input/output panels and QR
workspace still contribute at the gathered-slice memory scale.

\subsection{Overlapping communication and computation}
The communication overhead of our parallel schemes can be reduced by making some communication happen simultaneously with the computation.
Since we always divide the $l$ index into width-$w$ slices that are evaluated sequentially on each device, the gathering communications of the computed slices can overlap the computation of the next slices.

\subsection{Parallel performance benchmark and  validation}

Here we benchmark the parallelization algorithms on GPU clusters with the CUDA libraries for numerical linear algebra and the MPI and NCCL libraries for communications. Specifically, we use MareNostrum~5 nodes of the Barcelona Supercomputing Center, which have four NVIDIA H100 64-GB GPUs per node, NVLink all-to-all communication within a node, and relatively slower InfiniBand communication across nodes.  We work with Float64 data and set the local Hilbert-space dimension $d=2$ in the benchmark.

\begin{figure}[htbp]
\centering
\includegraphics[width=0.94\textwidth]{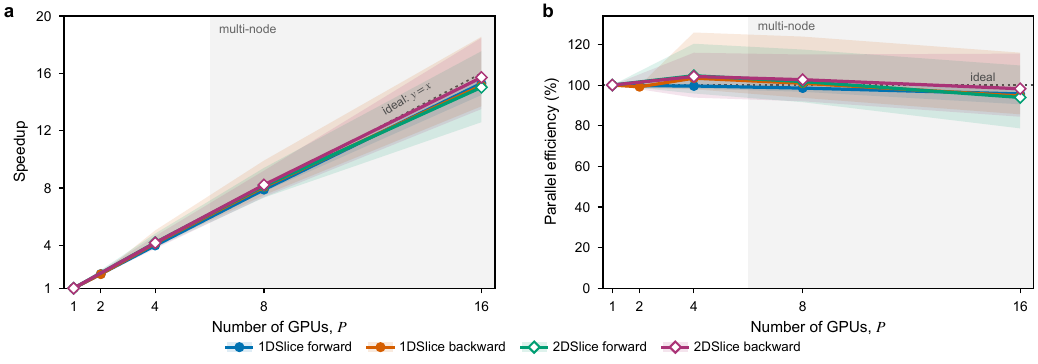}
\caption{\textbf{Strong scaling of FLmap and its VJP.}
Float64 measurements on NVIDIA H100 64-GB GPUs, four per node.
\textbf{a} Speedup relative to single-GPU run.
\textbf{b} Parallel efficiency. For 1DSlice (filled circles) and 2DSlice (open diamonds), and for forward and backward (VJP) calculations, we averaged speedup for 
$D=8,10,12$ and $\chi=512,1024$. Backward timings
include forward recomputation. Colored bands span the range from the minimum to maximum of choices of $D$ and $\chi$.  
For 2DSlice, we use $2\times2$,
$2\times4$, and $4\times4$ grids for the parallelization with  4, 8, and 16 GPUs.  Dotted lines indicate ``ideal'' speedup;
the light gray background marks the multi-node region ($P>4$).}
\label{fig:multigpu_scaling}
\end{figure}

In~\cref{fig:multigpu_scaling}, we show the strong-scaling performance with the number of GPUs for computing the dominant VUMPS environment tensor maps (forward pass) and the corresponding backward vector-Jacobian products (VJPs) used for gradient evaluation. The reported timings are the minimum over ten trials and are close to the average runtime observed in full optimization tasks. As discussed above, the environment tensor maps constitute the dominant computational cost for contracting the tensor network and evaluating observables, while the backward pass dominates gradient-based optimization. Let $T(P)$ denote the runtime for a given forward or backward calculation using $P$ GPUs. The strong-scaling results in~\cref{fig:multigpu_scaling} are presented in terms of the speedup $T(1)/T(P)$ [\cref{fig:multigpu_scaling}a]. We choose $D=8,10,12$ and $\chi=512,1024$ in the benchmark. The averaged speedup turns out to be very close to $P$. In other words, the parallel efficiency $T(1)/[T(P)P]$ is close to $100\%$ [\cref{fig:multigpu_scaling}b].

The speedup depends on the choice of $D$ and $\chi$, and the variation is shown as a colored band around the curves. In general, the more computationally demanding case is that with larger $\chi$ and $D$, where the efficiency increases. This is consistent with the fact that the communication data grow much slower in $\chi$ and $D$ than the computation cost.  However, this alone cannot explain efficiencies exceeding $100\%$, which is most notable for the memory-distributed 2DSlice scheme at larger $D$ and $\chi$. This is because reducing peak memory per GPU helps each GPU run closer to peak performance. (At $D=12$, the single-GPU run has nearly reached its memory limitation.) In our test, we adjust $l$-leg (sub) slicing width $w\in\{1,2,4,8,16\}$ for each run and choose its fastest timing. For different $w$, the timing can differ by as much as $10 \%$, affected by peak memory reduction, kernel efficiency, and/or communication overhead. 

In the test platform, the cross-node communication ($25$ GB/s card-card) is much slower than in-node NVLink (900 GB/s) communication, but it does not seem to be a bottleneck, as we see that the two-node run is nearly as efficient as the single-node run [\cref{fig:multigpu_scaling}]. For example, for $D = 12$ and $\chi=1024$, the 8-GPU 1DSlice FLmap forward pass takes $436$ ms, while the communication data divided by the peak communication speed is on the order of  $1 \ $ms for NVLink and $10 \ $ms for internode communication. This echoes our design principle of 2DSlice, for which the number of communication operations is kept small. As a good side for practical implementations,  the tensor parallelization for iPEPS-VUMPS is quite different from some other applications in the sense that super high-speed communication like NVLink is less demanded.

The benchmark in \cref{fig:multigpu_scaling} times single applications of the environment tensor maps. To verify that their parallel efficiency also determines that of the complete calculation, we time complete contraction (dominant tensor maps and TSQR/TSLQ canonicalizations) and energy evaluation along with the full-process reverse-mode differentiation. The results are compared with those of FLmap and its VJP in~\cref{tab:multigpu_e2e}. For 2DSlice, we use four GPUs as the reference because the energy-gradient evaluation at $D=12$, $\chi=1024$ does not fit on a single GPU. For both schemes, the overall efficiency closely follows that of the environment tensor maps; the parallel performance of the complete calculation is thus determined by that of the environment tensor maps. The small deviations come from the operations outside the environment maps, mainly the canonicalization, whose cost is subleading in $D$ as discussed above. 
The efficiency exceeding $100 \%$ of the $D=8$ energy-gradient evaluation is an effect of reduced peak memory due to memory distribution, same as our observation for FLmap efficiencies in~\cref{fig:multigpu_scaling}.

\begin{table}[htbp]
\caption{\textbf{Overall efficiency of the contraction-optimization algorithm.}
Parallel efficiency $\eta=P_0T(P_0)/[PT(P)]$ from $P_0$ to $P$ GPUs for single applications of FLmap and its VJP [fastest timing over $w$, as in \cref{fig:multigpu_scaling}] and for one complete evaluation of the energy and of the energy together with its gradient by AD. For 2DSlice, the $2\times2$ and $2\times4$ grids are used. Times $T(P_0)\to T(P)$ are also shown.
($^{a}$FLmap at $\chi=512$, the smallest $\chi$ in \cref{fig:multigpu_scaling}.)}
\label{tab:multigpu_e2e}
\begin{ruledtabular}
\begin{tabular}{cccccccc}
Scheme & $D$ & $\chi$ & $P_0\to P$ & FLmap & FLmap VJP & Energy & Energy and gradient\\
\hline
1DSlice & $12$ & $256$ & $1\to8$ & $96.1\%^{a}$ & $94.5\%^{a}$ & $89.7\%$ ($4.56\to0.64$~s) & $93.2\%$ ($23.7\to3.19$~s)\\
2DSlice & $8$ & $1024$ & $4\to8$ & $95.5\%$ & $97.9\%$ & $89.4\%$ ($12.4\to6.9$~s) & $107.8\%$ ($58.0\to26.9$~s)\\
2DSlice & $12$ & $1024$ & $4\to8$ & $97.8\%$ & $98.9\%$ & $96.0\%$ ($31.3\to16.3$~s) & $96.2\%$ ($148.5\to77.1$~s)\\
\end{tabular}
\end{ruledtabular}
\end{table}

Besides the profiling on up to 16 GPUs mentioned above, we also deploy the calculations on larger-scale parallelization on the H200 nodes of the Sofia cluster at the Flemish Supercomputer Center (VSC). We summarize the elapsed time record for a working example for the square lattice $J_1$-$J_2$ model in \cref{tab:multigpu_production}. 

\begin{table}[htbp]
\caption{\textbf{Large-scale calculations on Sofia H200 nodes.}
Full time (h:mm:ss) for optimizing the $J_1$-$J_2$ model using 2DSlice parallelization, and for an extremely high-precision VUMPS environment solve followed by energy calculation using 1DSlice parallelization.}
\label{tab:multigpu_production}
\begin{ruledtabular}
\begin{tabular}{lcccc}
Calculation & $D$ & $\chi$ & GPUs & Elapsed time\\
Optimization & 16 & 768 & 64 (2DSlice $8\!\times\!8$) & 8:48:03 \\
Observables & 16 & 1024 & 128 (1DSlice) & 0:15:58 

\end{tabular}
\end{ruledtabular}
\end{table}

\FloatBarrier

Most calculations of this work can be completed
within a few days on 16 GPUs. The tests on up to 128 GPUs verify that
parallel execution remains feasible at larger scale, in preparation for
more demanding applications.

\FloatBarrier

\end{document}